\documentclass[
preprint,
superscriptaddress, nofootinbib, aps,pra ]{revtex4}

\usepackage{braket}             % Allows Dirac notation
\usepackage{graphicx}           % Include figure files
\usepackage{bm}             % bold math
\usepackage{amsmath,amssymb,amsthm}
\usepackage[usenames, dvipsnames]{color}
\usepackage{xcolor}
\usepackage{hyperref}           % add hypertext capabilities
\hypersetup{
    colorlinks,
    linkcolor={blue!50!black},
    citecolor={blue!50!black},
    urlcolor={blue!80!black}
}
\usepackage{mathtools}
\usepackage{soul}               % for crossing things out
\usepackage{mathrsfs}
\usepackage{subfigure}
\usepackage{ulem}
\DeclareMathOperator{\tr}{tr}           % for trace

\newcommand{\nn}{\nonumber}     % for no numbering

\usepackage{colonequals} %Get a nice looking \ce
\newcommand{\ce}{\colonequals}

\begin{document}
\title{Entanglement harvesting for  Unruh-DeWitt detectors in circular motion}
\author{Jialin Zhang and Hongwei Yu\footnote{Corresponding author at hwyu@hunnu.edu.cn} }

\affiliation {Department of Physics and Synergetic Innovation Center for Quantum Effects and Applications, Hunan Normal University, Changsha, Hunan 410081, China}

\begin{abstract}
We study the properties of the transition probability and
entanglement harvesting phenomenon for circularly accelerated
detectors locally interacting with massless scalar fields. The
dependence of the transition probability on the parameters
associated with the circular motion is first analyzed in detail. By
a cross-comparison with the situation of the uniformly accelerated
motion, we obtain that the transition probability and the possible
thermalization behavior for detectors rotating  with an extremely
large circular radius are analogous to that for uniformly
accelerated detectors, but for a very small linear speed and a large
acceleration, the effective temperature which characterizes the
detectors' thermalization  in a finite duration is much lower than
that for uniformly accelerated detectors. We then focus on the
phenomenon of entanglement harvesting in two special situations of
circular  trajectories, i.e., the coaxial rotation and the
mutually perpendicular axial rotation by examining the concurrence
as the entanglement measure in detail. We find that when two circularly
accelerated detectors have  equivalent acceleration and size of
circular trajectory, the harvested entanglement  rapidly decays with
increasing acceleration or separation between two detectors. In
contrast with the situation of uniform acceleration, the angular
velocity would have significant impacts on entanglement harvesting.
Especially for those detectors circularly moving in different
directions, both the acceleration and trajectory radius play an
important inhibiting role in entanglement harvesting. When two
circularly accelerated detectors have different values of
acceleration or angular velocity, we find that the entanglement can
still be extracted by such detectors, even in the situation that one
detector is at rest and the other is in a circular motion.

\end{abstract}
\maketitle
\section{Introduction}
 Quantum entanglement has been widely known as a key
physical resource for performing some tasks in quantum
information~\cite{Plenio:1998,Horodecki:2001}, and much progress has
been made in understanding the features of entanglement in various aspects.  For a two-atom system coupled with a  common bath
or environment, it has
been found, in the framework of open quantum systems,  that the generation  of entanglement is likely to happen
in certain
circumstances~\cite{Braun:2002,MSkim:2002,Schneider:2002,Basharov:2002,Jakobczyk:2002,Reznik:2003,Benatti:2003,Ficek:2003},
and some other general entanglement dynamical behaviors such as the phenomenon  of
entanglement sudden death~\cite{TYu:2004,Eberly:2007} or
entanglement revival~\cite{Ficek:2006} may also emerge. Recently, a more operational approach to studying quantum
field entanglement has arisen from Valentini's pioneer
work~\cite{Valentini:1991}(and later revisited by
Reznik~\cite{Reznik:2003}). It has been argued that a pair of
initially uncorrelated atoms can become entangled via  locally
interacting with vacuum fields, even if they remain spacelike
separated. Such proposed process has been extensively investigated
in various circumstances involving the curved spacetime of a black
hole~\cite{Ver
Steeg:2009,Olson:2011,BLHu:2012,Pozas-Kerstjens:2015,EDU:2016-1,EDU:2016-2,Pozas-Kerstjens:2016,Zhjl:2018,Zhjl:2019,Ng:2018,Ng:2018-2},
which has  now  become recognized as the entanglement harvesting
protocol~\cite{Salton:2015}. Since the entanglement harvesting
phenomenon is sensitive to the curvature of spacetime, it may
be used to discern the  structure of
spacetime~\cite{EDU:2016-1} and distinguish a thermal background
from the Hawking radiation background of an expanding
universe~\cite{Ver Steeg:2009}.

On the other hand, the Unruh effect attests that accelerated detectors in
Minkowski vacuum will observe a thermal radiation spectrum of
particles~\cite{Unruh:1976}, which is closely related to  the Hawking radiation,
one of the most striking predications in quantum field
theory in curved spacetime~\cite{Birrell:1984}. So, the Unruh effect is believed to offer a
promising way to understanding other phenomena such as Hawking radiation
of black holes and the thermal emission from cosmological
horizons~\cite{Birrell:1984,Crispino:2008}. Nevertheless,  a direct
test of the Unruh effect  would  require some extreme physical
conditions which are currently inaccessible in laboratory. However,
a lot of effort has been made in exploring it through
different means, including some novel proposals for
experiments~\cite{Crispino:2008} and other potentially measurable
related quantum phenomena such as the geometric
phase~\cite{EDU:2011,Hu:2012,Zhjl:2016}, the Lamb
shift~\cite{Audretsch :1995,Passante:1998,Rizzuto:2009,Zhu:2010},
quantum
entanglement~\cite{Benatti:2004,Zhjl:2007,Landulfo:2009,Doukas:2010,Ostapchuk:2012,Hu:2015,Cheng:2018,Koga:2019,She:2019}.
More recently, it has been argued that there is a counter-intuitive
phenomena (so-called anti-Unruh phenomena) that particle  detectors
can click less often or even cool down with the increase of  the
acceleration under certain conditions~\cite{Brenna:2016,Garay:2016}.
Within certain parameter regime, the anti-Unruh effect  may possibly
be viewed as an amplification mechanism for quantum
entanglement~\cite{Li:2018}.

Another interesting issue concerning the Unruh effect is the entanglement harvesting for  accelerated detectors. It is worth noting that although this issue  has recently  been discussed in detail for the case of two  linearly accelerated  detectors in
Refs.~\cite{Salton:2015,Koga:2019}, relatively
little is known about entanglement harvesting for a pair of
circularly accelerated detectors.  Actually, as opposed to  the
situation of constant linear acceleration, the circular  motion seems
more interesting since it is  easier to achieve, in a circular motion, the necessary
large acceleration needed in the experimental verification of the
Unruh effect~\cite{Bell:1983}.  Recently,  discussions  on the
entanglement dynamics of circularly accelerated atoms coupled with
the electromagnetic vacuum have been performed  in Ref.~\cite{She:2019},
but these results are limited to the Born-Markov approximation,
requiring the pair of  atoms have the same angular velocity and
acceleration.

In this paper we will perform a  more general study of the
entanglement harvesting phenomenon of two circularly accelerated
detectors, relaxing the limiting condition of the same angular
velocity and acceleration. For simplicity, we will employ  the
well-known Unruh-DeWitt (UDW) model to depict the particle detector
interacting with vacuum quantum fields~\cite{DeWitt:1979}.  The
paper is organized as follows. First, some basic formulae for the
UDW detectors locally interacting with vacuum scalar fields are
reviewed with the help of the entanglement harvesting protocol. In
Sec. III, we will study the influence of the parameters concerned
with the circular motion on the transition probability, such as
the acceleration and circular trajectory radius. In addition, we
also  allow for a cross-comparison of the transition probabilities
between the situation of circular acceleration and  linear
acceleration. In section IV, we will consider the entanglement
harvesting phenomenon for a pair of circularly accelerating
detectors along general circular trajectories, involving the coaxial
and non-coaxial rotations. Some necessary numerical evaluation will
be called for in the investigation. Finally, we conclude  the paper
with a summary in Sec.V. Throughout this paper  the natural units
$\hbar = c = 1$ are adopted for convenience.

\section{The basic formalism for
entanglement harvesting protocol}

In this section, we will introduce the  description of a two-level
atom interacting locally with a quantum scalar filed.    We also review
the derivation of basic formulas  in entanglement harvesting
protocol. Without loss of generality,  two  such atoms (labeled by
$A$ and $B$ ) can be modeled with the UDW detectors. Now supposing that
the spacetime trajectory of a detector is parameterized in terms of
its proper time, then the interacting Hamiltonian for such a detector
locally coupling with a real scalar field $\phi(x)$ has the
following form in the interaction picture
\begin{equation} \label{Int2}
H_D(\tau_D)=\lambda\chi_D(\tau_D)\Big(e^{i\Omega_D\tau_D}\sigma^++e^{-i\Omega_D\tau_D}\sigma^-\Big)\otimes\phi\big[x_D(\tau_D)\big]\;,~~D\in\{A,B\}
\end{equation}
where $\lambda$ is the coupling strength  which is assume to be weak,
$\chi_D(\tau_D):=e^{-\tau_D^2/(2\sigma_D^2)}$ is a Gaussian
switching function controlling the duration of interaction via the
parameter $\sigma_D$, and $\sigma^\pm$ denote the ladder operators
acting on the Hilbert space of the detector. Particularly, for a
two-level atom system with an energy gap $\Omega_D$,  we have
$\sigma^+=\ket{1}_D\bra{0}_D$ and  $\sigma^-=\ket{0}_D\bra{1}_D$
with $\ket{0}_D$  and $\ket{1}_D$  respectively denoting the ground
and excited states. Here, the subscript $D$ specifies which UDW
detector we are considering.

Suppose two such detectors $A$ and $B$ to be initially  in their
ground states, coupled with the scalar field in vacuum state
$\ket{0}$, then the initial joint state can be written as
$\ket{\Psi}=\ket{0}_A\ket{0}_B\ket{0}$. For simplicity, we assume
that all detectors have an identical energy gap
($\Omega_D=\Omega,~~D\in\{A,B\}$) and switching parameter
($\sigma_D=\sigma,~~D\in\{A,B\}$) in their own rest frame. Governed by  Hamiltonian Eq.~(\ref{Int2}), the composite
system (two detectors plus the field) will undergo the unitary
evolution  with the corresponding operator satisfying
\begin{equation}
U:={\cal{T}} \exp\Big[-i\int{dt}\Big(\frac{d\tau_A}{dt}H_A(\tau_A)+
\frac{d\tau_B}{dt}{H_B}(\tau_B)\Big)\Big]\;,
\end{equation}
where ${\cal{T}}$ denotes the time ordering operator. After some
manipulations based on the perturbation theory, the finial state of
the two detectors can be obtained by tracing out the field degrees
of freedom~\cite{EDU:2016-1,Zhjl:2018,Zhjl:2019}
\begin{align}\label{rhoAB}
\rho_{AB}:&=\tr_{\phi}\big(U\ket{\Psi}\bra{\Psi}U^+\big)\nonumber\\
&=\begin{pmatrix}
1-P_A-P_B & 0 & 0 & X \\
0 & P_B & C & 0 \\
0 & C^* & P_A & 0 \\
X^* & 0 & 0 & 0 \\
\end{pmatrix}+{\mathcal{O}}(\lambda^4)\;,
\end{align}
where the basis $\{\ket{0}_A\ket{0}_B,\ket{0}_A\ket{1}_B,\ket{1}_A\ket{0}_B,\ket{1}_A\ket{1}_B\}$ has been used.
Here, the corresponding parameters in the reduced density matrix $\rho_{AB}$ read
\begin{equation}\label{probty}
P_D:=\lambda^2\iint{d\tau_D}{d\tau_D'}\chi_D(\tau_D)\chi_D(\tau_D')e^{-i\Omega(\tau_D-\tau_D')}W(x_D(\tau_D),x_D(\tau_D'))\;\;\;\;
D\in\{A,B\}\;,
\end{equation}
\begin{align}
C:=&\lambda^2 \iint dt  dt' \,  \frac{\partial\tau_B}{\partial{t}} \frac{\partial\tau_A}{\partial{t'}} \chi_B(\tau_B(t))  \chi_A(\tau_A(t')) e^{i \left[ \Omega\tau_B(t)-\Omega \tau_A(t'\right)]} W\!\left(x_A(t') , x_B(t)\right)\;,
\end{align}
and
\begin{align}\label{defX}
X&\ce-\lambda^2  \iint_{t>t'} dt  dt'  \bigg[
\frac{\partial\tau_B}{\partial{t}} \frac{\partial\tau_A}{\partial{t'}} \chi_B(\tau_B(t)) \chi_A(\tau_A(t'))   e^{-i\left[\Omega  \tau_B(t)+\Omega \tau_A(t')\right]} W\!\left(x_A(t'), x_B(t)\right) \nn \\
 & \quad \qquad \qquad \qquad \quad  + \frac{\partial\tau_A}{\partial{t}}\frac{\partial\tau_B}{\partial{t'}} \chi_A(\tau_A(t)) \chi_B(\tau_B(t'))  e^{-i\left[ \Omega \tau_A(t)+ \Omega \tau_B(t')\right]} W\!\left(x_B(t'),x_A(t) \right)  \bigg],
\end{align}
where $W(x,x'):=\bra{0}\phi(x)\phi(x')\ket{0}$ is the Wightman
function associated with the scalar field. In fact, $P_D$ denotes
the probability that a detector has transitioned from its ground state to
the excited state due to its interaction with the
field~\cite{EDU:2016-1}, and $X$ represents the non-local
correlations between two detectors~\cite{Zhjl:2019}.

According to the entanglement harvesting
protocol~\cite{Salton:2015}, we can employ the  concurrence as a
measure of entanglement~\cite{Wootters:1998}, which can be
evaluated straightforwardly from the $X$-like density matrix given
in Eq.~(\ref{rhoAB}) to yield the concurrence~\cite{EDU:2016-1}
\begin{equation}\label{con1}
\mathcal{C}(\rho_{AB})=2\max\big\{0,|X|-\sqrt{P_AP_B}\big\}+{\mathcal{O}}(\lambda^4)\;.
\end{equation}
Obviously, such entanglement measure is a competition between
the  non-local correlation  $X$ and the transition probabilities,
which in general is determined by the Wightman function of scalar
fields. For the purpose of studying  the entanglement harvesting
phenomenon for circularly accelerated detectors, it is convenient to
give the Wightman function and  first examine  the behavior of transition
probabilities  in the detector's frame.

\section{The transition probabilities of circularly accelerated UDW detectors}
 In a 4-dimensional Minkowski spacetime,  the Wightman function for massless scalar fields
 can be given in the popular ``$i\epsilon$" representation \cite{Birrell:1984}
 \begin{equation}\label{wightman1}
W(x,x')=-\frac{1}{4\pi^2}\frac{1}{(t-t'-i\epsilon)^2-|{\bf {x}}-{\bf
{x'}}|^2}\;.
 \end{equation}
The spacetime trajectory of circular motion can be
parameterized by detector's proper time
$\tau_D$~\cite{She:2019,Doukas:2010,Kim:1987}
\begin{equation}\label{traj1}
x_D:=\{t=\gamma_D\tau_D\;, ~~x=R_D\cos(\omega_D\gamma_D\tau_D)\;,
~~y=R_D\sin(\omega_D\gamma_D\tau_D)\;,~~z={\rm{const}} \}\;,
\end{equation}
where $R_D$ represents the radius of the circular trajectory in a
plane parallel to $xy$-plane, $\omega_D$ is the angular velocity
which can be either positive or negative in circular motion,
 and $\gamma_D=1/\sqrt{1-R_D^2\omega_D^2}$ denotes the Lorentz factor.  In the detector's frame, the magnitude of acceleration satisfies $a_D=\gamma_D^2\omega_D^2R_D=\gamma_D^2v_D^2/R_D$,
with the magnitude of linear velocity  obeying
$v_D=|\omega_D|R_D<1$. Note that $\omega_D$, $R_D$, $a_D$ and $v_D$
are not completely independent motion parameters,  only two
 of them are actually independent. Substitute the trajectory~(\ref{traj1}) into Eq.~(\ref{wightman1}),
we can get the Wightman function
\begin{equation}\label{wightman2}
W(\tau_D,\tau'_D)=-\frac{1}{4\pi^2}\frac{1}{(\gamma_D\Delta\tau-i\epsilon)^2-4R_D^2\sin^2(\gamma_D\omega_D\Delta\tau/2)}\;
\end{equation}
with $\Delta\tau=\tau_D-\tau'_D$.

 As we can see from
Eq.~(\ref{wightman2}) that the corresponding Wightman function does
not satisfy the Kubo-Martin-Schwinger (KMS) condition~\cite{Kubo:1957,Martin:1959}, i.e., we can not find a
nonzero $T_{\rm{KMS}}$ to ensure that the Wightman function obey
the relation:
\begin{equation}
W(\tau_D-i/T_{\rm{KMS}},\tau'_D)=W(\tau'_D,\tau_D)\;.
\end{equation}
Therefore, for the  circularly accelerated motion  there is
no well-defined  KMS temperature  in quantum field theory, which is
quite different from the situation of a linearly uniformly
accelerated detector with a KMS temperature proportional to the
magnitude of acceleration~\cite{Unruh:1976}.

Substituting Eq.~(\ref{wightman2}) into Eq.~(\ref{probty}), we find
that the transition probability  (see Appendix~{\ref{Derivation-PD}}
for detail)
\begin{align}\label{PAPB}
P_D=K_D\int_{0}^{\infty}dx\frac{\cos(
x\beta)e^{-x^2\alpha}(x^2-\sin^2x)}{x^2(x^2-v_D^2\sin^2x)}+\frac{\lambda^2}{4\pi}\Big[e^{-\Omega^2\sigma^2}-\sqrt{\pi}\Omega
\sigma {\rm{Erfc}}\big(\Omega\sigma\big)\Big]
\end{align}
where
\begin{equation}
\alpha=\frac{1}{\sigma^2\omega_D^2\gamma_D^2}=\frac{R_D}{a_D\sigma^2}\;,~~
\beta=\frac{2\Omega}{\gamma_D|\omega_D|}\;,~~K_D=\frac{\lambda^2v_D^2\gamma_D|\omega_D|\sigma}{4\pi^{3/2}}
=\frac{\lambda^2v_Da_D\sigma}{4\pi^{3/2}\gamma_D}\;,
\end{equation}
and ${\rm{Erfc}}(x)$ is the complementary error function, satisfying
the identity  ${\rm{Erfc}}(x)=1-{\rm{Erf}}(x)$. Although the first
term in Eq.~(\ref{PAPB}) is a regular integration, it is a messy
task to get a simple analytical result and some numerical
evaluations are needed~\cite{Doukas:2010}. However, for some certain
extreme cases,  approximate results can be obtained directly. For
example,  for  an extremely large acceleration  with high speed (i.e., $a_D\sigma\gg\gamma_D\gg1$ and $\left|\Omega\right|/a_D\ll1$), $P_D\approx{a_D}\sigma\lambda^2/(8\sqrt{3\pi})$ (
see Appendix~\ref{approx} for more details), while for a small
acceleration with high speed or extremely large radius (i.e.,
$\gamma_D\gg a_D\sigma$, $\gamma_D\gg1\gg{a}_D|\Omega|\sigma^2$), through a saddle point
approximation~\cite{Nambu:2013}, we can obtain ( see
Appendix~\ref{approx})
\begin{equation}
P_D\approx
\frac{a_D^2\sigma^2\lambda^2e^{-\sigma^2\Omega^2}}{24\pi}+\frac{\lambda^2}{4\pi}\Big[e^{-\Omega^2\sigma^2}-\sqrt{\pi}\Omega
\sigma {\rm{Erfc}}\big(\Omega\sigma\big)\Big]\;.
\end{equation}
Particularly, in the limit of $v_D\rightarrow0$, the first term of
Eq.~(\ref{PAPB}) is vanishing and the second term is just the
transition probability of a rest detector with a Gaussian switching
function, which is completely consistent with the result in
Refs.~\cite{EDU:2016-1,Nambu:2013}.

\begin{figure}[!ht]
\centering{
\includegraphics[width=0.8\textwidth]{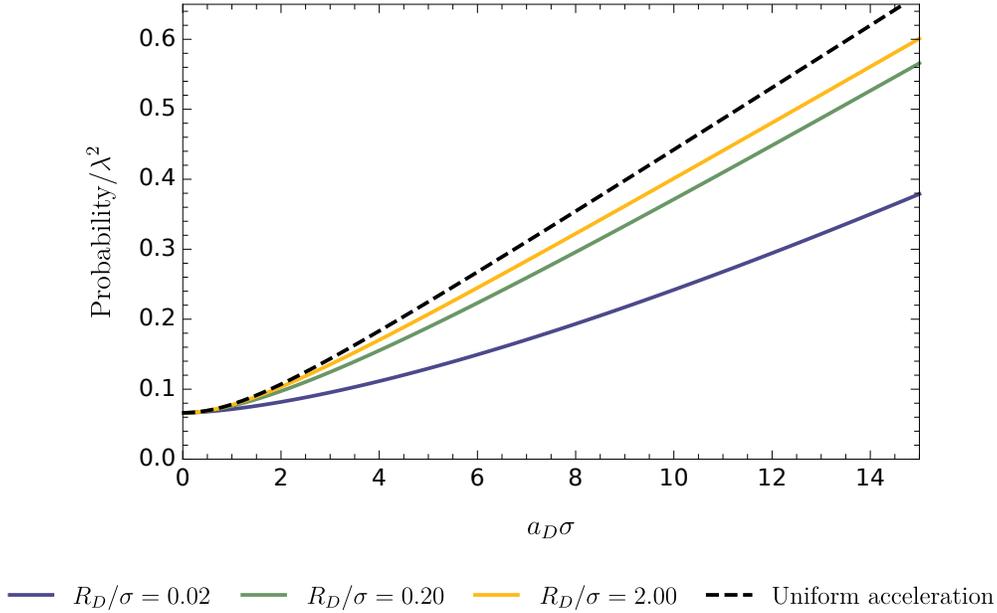}}
\caption{ The transition probability of detector $D$  is plotted as a function of  acceleration (denoted by dimensionless quantity $a_D\sigma$ for convenience) with fixed $\Omega\sigma=0.10$. Here, the circularly accelerated situation is depicted by the solid lines  and the dashed line represents the  uniformly accelerated situation.}\label{PAPBvsA}
\end{figure}

\begin{figure}[!ht]
\centering{
\includegraphics[width=0.8\textwidth]{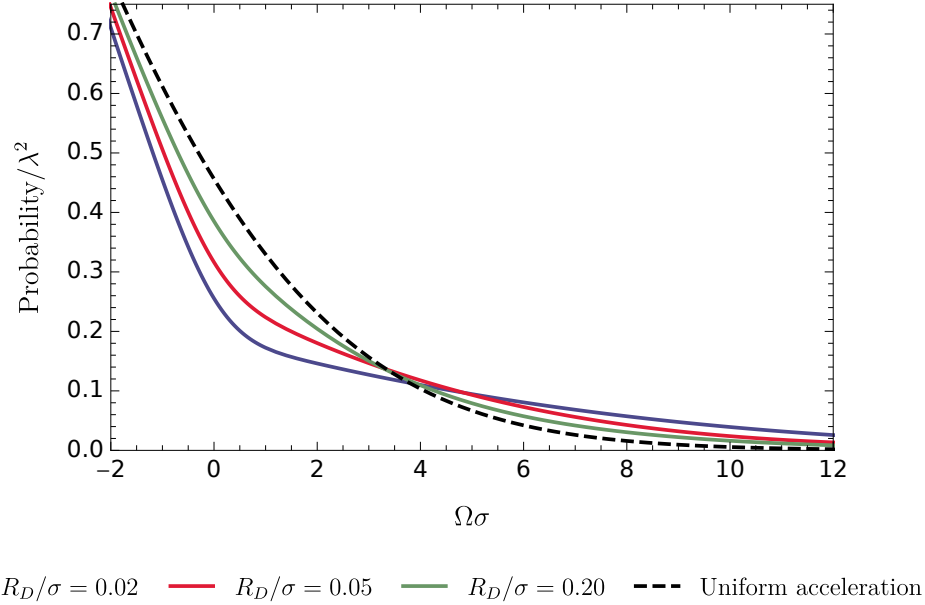}}
\caption{The transition probability is plotted as a function of its
energy gap $\Omega\sigma$ with fixed $a_D\sigma=10$. Here, the solid
lines denote the  transition probability for a detector rotating along the circular trajectories of different radii. It is worth pointing
out the negative (positive) energy gaps correspond to that detector
prepared in its excited (ground) state prior to interacting with the
filed.}\label{PAPBvsomega}
\end{figure}

\begin{figure}[!ht]
\centering{
\includegraphics[width=0.75\textwidth]{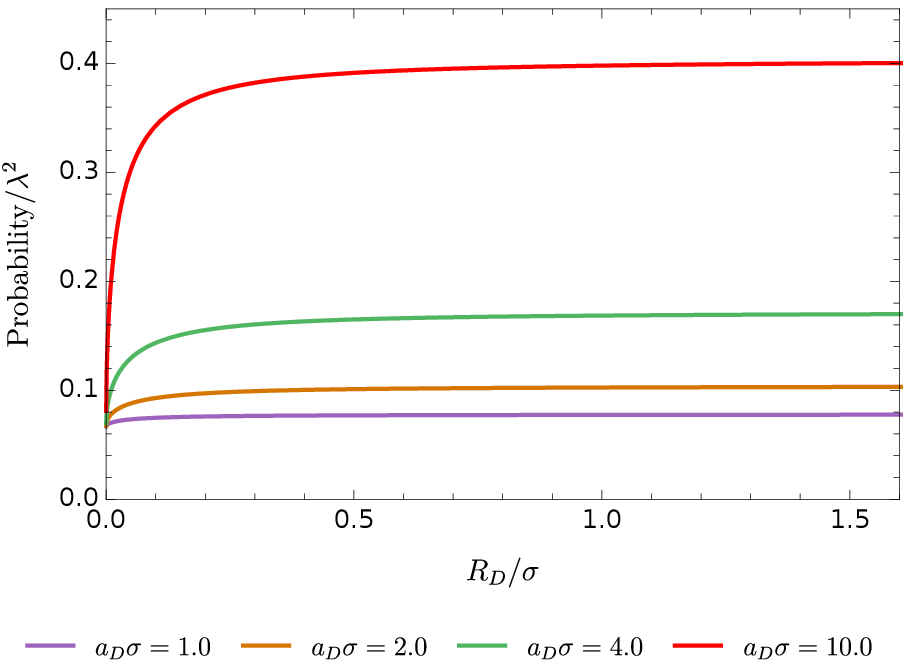}}
\caption{The transition probability of detector $D$ in circular
motion vs its radius of a circle  with the relevant energy gap  fixed as
$\Omega\sigma=0.10$.}\label{PAPBvsR}
\end{figure}

In order to understand how the transition probability depends on the
acceleration and other parameters in circular motion,
we illustrate the  detailed behavior of the transition probability
in Figs.~(\ref{PAPBvsA}-\ref{PAPBvsR}). Here, throughout all the following plots  the relevant physical quantities
are adapted by the corresponding dimensionless ones in the unit  of $\sigma$.

%%%%%%%%%%%%%%%%%%%%%%%%%%


%%%%%%%%%%%%%%%%%%
 To allow for a cross-comparison of  transition probabilities, here we
 have considered the following world line for a uniformly accelerated
detector~\cite{Birrell:1984,Crispino:2008,Rizzuto:2009,Salton:2015}
\begin{equation}\label{un-traj}
x_D:=\{t=a_D^{-1}\sinh(a_D\tau)\;,~~x=a_D^{-1}\cosh(a_D\tau)\;,~~y={\rm{const}}\;,~~z={\rm{const}}\}\;,
\end{equation}
where the magnitude of the linear constant acceleration is still
denoted by  $a_D$. Similarly, the numerical evaluation of transition
probabilities can also be carried out using Eq.~(\ref{PApB-UA}) in
Appendix~\ref{Derivation-PD} (for related discussions see
Ref.~\cite{Salton:2015}). As shown in Fig.~(\ref{PAPBvsA}), the
corresponding transition probability is a generally increasing
function  of acceleration irrespective of the circular or linear
motion,  and  there does not seem to be  anti-Unruh effect for the
circularly accelerated detectors in Minkowski spacetime in terms of
the transition probability.  More interestingly,  the transition
probability for  the linear uniform acceleration is more sensitive to the
increasing acceleration, while
 for the circular motion,  the smaller the trajectory radius is, the
less variation the transition probability entails.

In Fig.~(\ref{PAPBvsomega}), we explore the influence of the energy gap
$\Omega$ on the transition probability. It is easy to see that  the
transition probability is a monotonically decreasing function  of
$\Omega\sigma$. Especially  for positive $\Omega\sigma$,
corresponding to the initial ground state, the transition to excited
state hardly happens for a big energy  gap. This is
consistent with our intuition that the larger the energy gap the harder the transition is to happen, which also follows  straightforwardly from a mathematical examination of the general form of transition probability
Eq.~(\ref{PAPB}).
 It is worth pointing out that the larger the trajectory radius $R_D$ is, the faster
the transition probability decays with increasing energy gap.
These properties can be understood from the corresponding integrand
in Eq.~(\ref{PAPB}). For a fixed acceleration,
$\beta=2\Omega\sqrt{R_D/a_D}$ is associated to the highly
oscillatory part of Eq.~(\ref{PAPB}). As the energy gap increases, a
large trajectory radius will render the  value of parameter $\beta$
much larger than 1, thus the integration part of Eq.~(\ref{PAPB})
becomes vanishingly small as a result of the rapid oscillation of
the cosine function.

To examine the influence of  the circular trajectory radius  on the transition
probability, we plot the transition probability as a function of $R_D/\sigma$  in Fig.~(\ref{PAPBvsR}). We find that the
transition probability is a generally increasing function of the radius for a fixed $a_D\sigma$,  but the increased amount will be quite small as $R_D/\sigma$ becomes large. It is worth pointing out that the limit of $R_D=0$ just corresponds to
the situation of a rest point detector in Minkowski spacetime, and
the transition probability is determined by the second term of Eq.~(\ref{PAPB}) which is independent of the acceleration.

In order to get a better understanding of the possible thermalization
process, it is convenient  to define an effective temperature called
$T_{{\rm{EDR}}}$ by utilizing the excitation to de-excitation ratio
(EDR) of the detector~\cite{Fewster:2016}, that is
\begin{equation}\label{Tedgs}
T_{{\rm{EDR}}}=-\frac{\Omega}{\log{\mathcal{R}}}\;,
\end{equation}
where ${\mathcal{R}}={\mathcal{F}}(\Omega)/{\mathcal{F}}(-\Omega)$
represents the EDR ratio. With the Gaussian switching
function, the corresponding response function
${\mathcal{F}}(\Omega)$ can be written in terms of the transition
probability as
\begin{equation}
{\mathcal{F}}:=\frac{P_D}{\lambda^2\sigma}\;.
\end{equation}

In general, the EDR temperature defined in Eq.~(\ref{Tedgs}) is
complicated and dependent on the parameters $a_D$, $\Omega$, $v_D$.
However, in the limit of an infinite interaction time and an
extremely high speed ($\sigma\rightarrow\infty\;,v_D\rightarrow1$),
the EDR temperature for circular acceleration is approximated to
$a_D/2\sqrt{3}$ for $a_D\ll|\Omega|$ (see Appendix~\ref{approx}),
which is higher than the EDR temperature $a_D/(2\pi)$ for linear
uniform acceleration~\cite{Bell:1983}.

\begin{figure}[!ht]
\centering{
\includegraphics[width=0.8\textwidth]{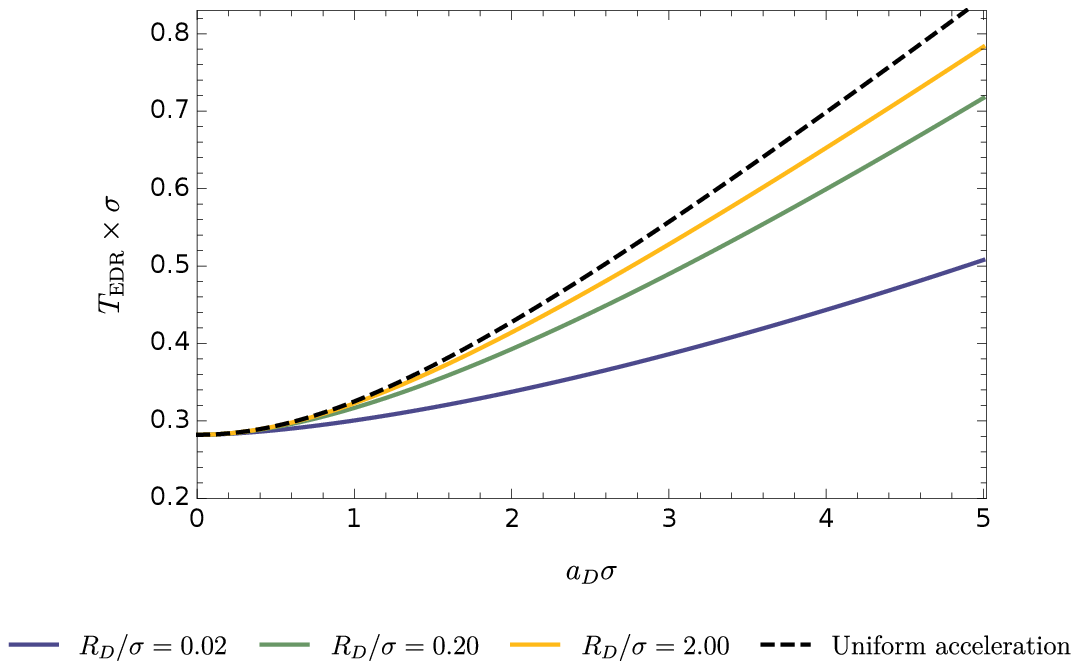}}
\caption{The effective temperature $T_{{\rm{EDR}}}$ associated with
the transition probability  is plot as a function of the magnitude
of acceleration. Here, we have set $\Omega\sigma=0.10$ for both
circularly  and  uniformly accelerated motion.}\label{PAPBvsTed}
\end{figure}

 To illustrate the general thermalization process in a finite duration time,  we have
plotted how the EDR temperature depends on the acceleration  at
various trajectory radii in Fig.~(\ref{PAPBvsTed}). It is easy to
find that for a finite duration time the EDR temperature is an
increasing function of acceleration, but the effective temperature
for circular acceleration is lower than that for uniform
acceleration when the energy gap is not too big, i.e., in comparison
with circularly accelerated detectors, the uniformly accelerated
detector would observe stronger thermal-like noise at the same
magnitude of acceleration. Particularly, for a vanishingly small
speed and a large acceleration with a not-extremely small energy
gap($1\gg{v_D}\;,a_D\sigma\gg|\Omega|\sigma>1\;$), the EDR
temperature for circular motion, according to Eq.~(\ref{PAPB}) and
Eq.~(\ref{Tedgs}),  approximately satisfies a simple relation:
$T_{{\rm{EDR}}}\approx a_D v_D \sqrt{1-v_D^2}/6$ (see
Appendix~\ref{approx}), which is much lower than the EDR temperature
$T_{{\rm{EDR}}}\approx {a_D}/(2\pi)$ for large linear uniform
acceleration ($a_D\sigma\gg|\Omega|\sigma>1$).

\section{Entanglement harvesting with UDW detectors in the circular motion}
\begin{figure*}[!ht]
\subfigure[]{\label{twoorbit11}
\includegraphics[width=0.45\textwidth]{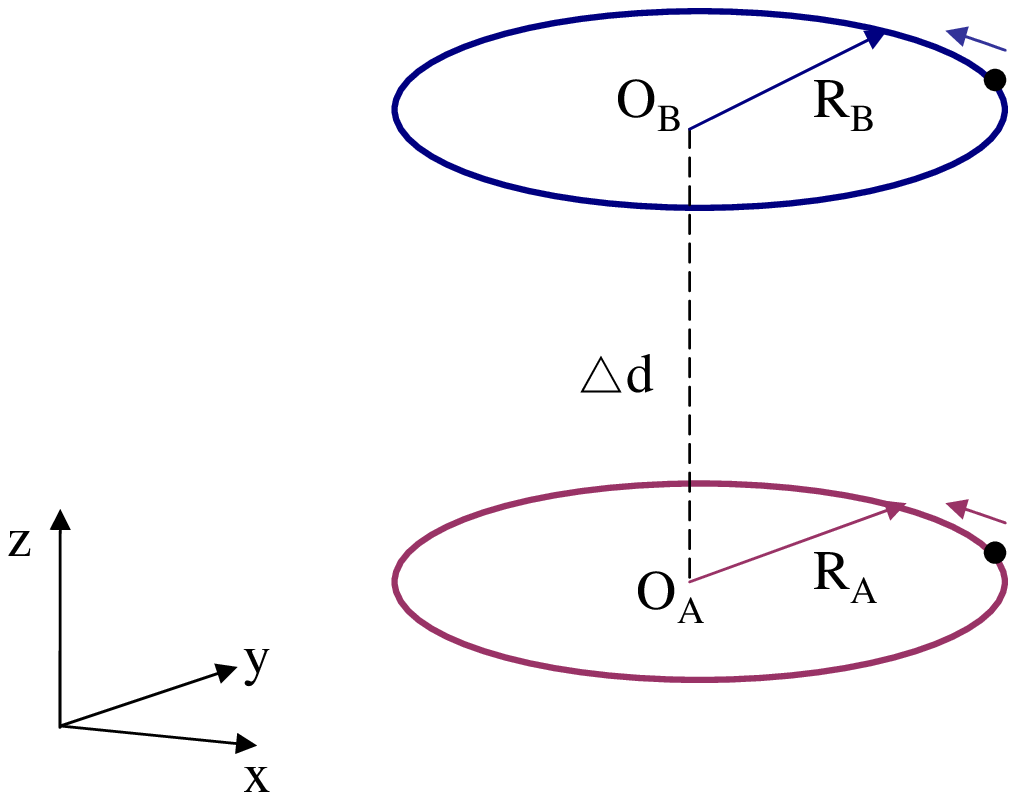}}\quad\subfigure[]{\label{twoorbit22}
\includegraphics[width=0.45\textwidth]{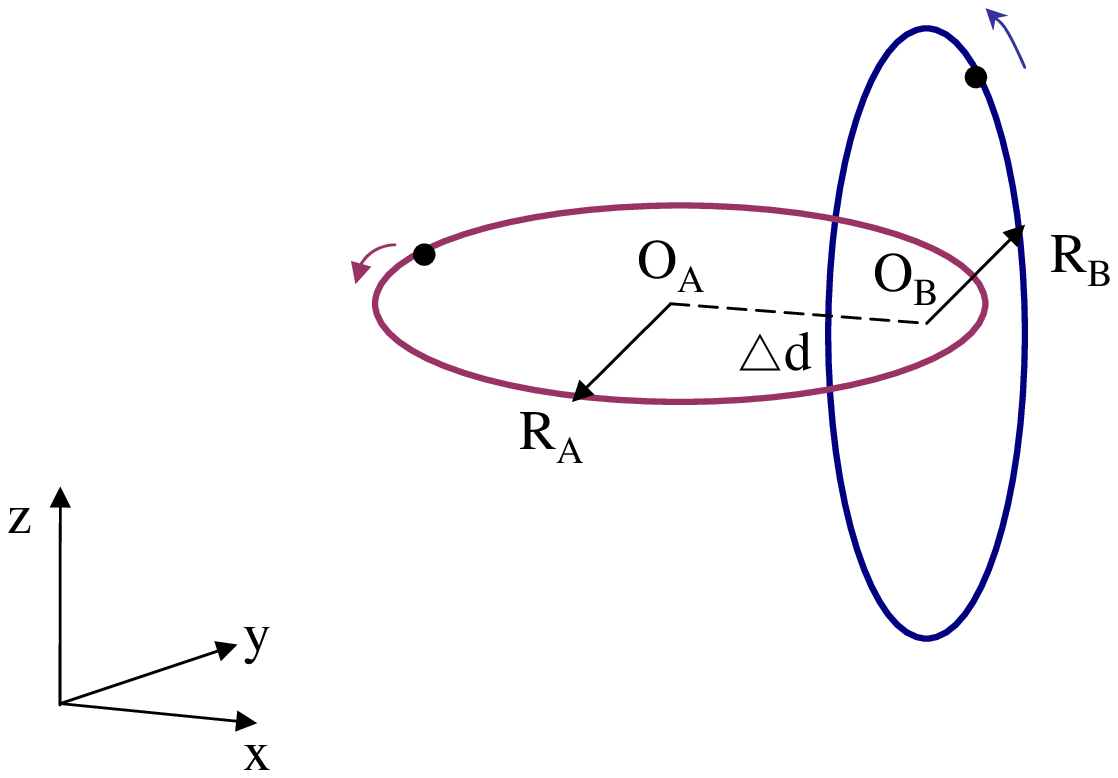}}
\caption{The circular trajectories for two UDW detectors $A$ and $B$
in flat spacetime. In (a), $z-$axis is the  common rotational axis
of such two detectors , and in (b) the corresponding rotational axes  are
mutually perpendicular.}\label{twoorbit}
\end{figure*}

We now explore the entanglement harvesting phenomenon of two
circularly accelerated detectors. For simplicity, we mainly focus on
the spacetime trajectories of the detectors in two special cases:
coaxial rotation and mutually perpendicular axial rotation (see
Fig.~(\ref{twoorbit})). Once having specified the trajectories, the
concurrence can be calculated via the afore-given formulas.

\subsection{ the situation of coaxial rotations}
Suppose that detectors $A$ and $B$ with angular velocities
$\omega_A$ and $\omega_B$ rotate  around the $z$-axis with the radii
$R_A$ and $R_B$.  For the circular motion~(\ref{traj1}),  the
spacetime trajectories of the two detectors  can be parameterized
respectively by their proper times $\tau_A$ and $\tau_B$
\begin{align}\label{traj2}
&x_A:=\{t=\tau_A\gamma_A\;, x=R_A\cos(\omega_A\tau_A\gamma_A)\;,
y=R_A\sin(\omega_A\tau_A\gamma_A)\;,z=0\}\;,
\nonumber\\
&x_B:=\{t=\tau_B\gamma_B\;, x=R_B\cos(\omega_B\tau_B\gamma_B)\;,
y=R_B\sin(\omega_B\tau_B\gamma_B)\;,z=\Delta{d}\}\;.
\end{align}
Here, $\gamma_A$ and $\gamma_B$ are corresponding Lorentz factors of
detectors $A$ and $B$ respectively. The parameter $\Delta{d}$ is the
separation between the two  centers.

With the spacetime trajectories~(\ref{traj2}), it is easy to find
that the transition probabilities of the detectors can be
straightforwardly calculated by using Eq.~(\ref{PAPB}).  The quantity  $X$
representing the non-local correlations  can be obtained by substituting
 Eq.~(\ref{traj2}) and Eq.~(\ref{wightman1}) into Eq.~(\ref{defX}). For convenience, here we use  $X^{\parallel}$ to stand for $X$  in the case where  the trajectories of two detectors are in parallel orbital planes described by
Eq.~(\ref{traj2}),  thus  result of $X^{\parallel}$  can  be generally written, after some algebraic
 manipulations, as
\begin{align}\label{Xint1-1}
X^{\parallel}&=-
\frac{\lambda^2\sigma^2}{4\pi^2\gamma_A\gamma_B}
\int_{-\infty}^{\infty} d\tilde{u} \int_{0}^{\infty} d\tilde{s}
\,\bigg\{
\exp\Big[\frac{-\gamma_A^2\tilde{u}^2-\gamma_B^2(\tilde{s}-\tilde{u})^2}{2\gamma_A^2\gamma_B^2}\Big]
\exp\Big[\frac{i(\tilde{s}-\tilde{u})\sigma\Omega}{\gamma_A}-\frac{i\tilde{u}\sigma\Omega}{\gamma_B}\Big] \nonumber\\
&
\times{f_{AB}}(\tilde{u},\tilde{s})+\exp\Big[\frac{-\gamma_B^2\tilde{u}^2-\gamma_A^2(\tilde{s}-\tilde{u})^2}{2\gamma_A^2\gamma_B^2}\Big]
\exp\Big[\frac{i(\tilde{s}-\tilde{u})\sigma\Omega}{\gamma_B}-\frac{i\tilde{u}\sigma\Omega}{\gamma_A}\Big]{f_{BA}}(\tilde{u},\tilde{s})\bigg\}\;,
\end{align}
where  the auxiliary functions read
\begin{equation}
f_{AB}(\tilde{u},\tilde{s})=\Big[(\Delta{d})^2+R_A^2+R_B^2-2R_AR_B\cos\big({\tilde{u}
\omega_A\sigma}-{\tilde{u}\omega_B\sigma-\tilde{s}\omega_A\sigma}\big)-\sigma^2(\tilde{s}+i\epsilon)^2\Big]^{-1}\;,
\end{equation}
\begin{equation}
f_{BA}(\tilde{u},\tilde{s})=\Big[(\Delta{d})^2+R_A^2+R_B^2-2R_AR_B\cos\big({\tilde{u}
\omega_A\sigma}-{\tilde{u}\omega_B\sigma+\tilde{s}\omega_B\sigma}\big)-\sigma^2(\tilde{s}+i\epsilon)^2\Big]^{-1}\;.
\end{equation}

 When such two detectors are completely synchronously rotating
around $z$-axis, i.e., $\omega_A=\omega_B=\omega$,
Eq.~(\ref{Xint1-1}) can  be further simplified to a one-dimensional
integral
\begin{align}\label{Xint1-2}
X^{\parallel}=&-
\frac{\lambda^2\sigma^2}{\pi^{3/2}\sqrt{2(\gamma_A^2+\gamma_B^2)}}\exp\Big[\frac{-\sigma^2\Omega^2(\gamma_A+\gamma_B)^2}{2(\gamma_A^2+\gamma_B^2)}\Big]
\int_{0}^{\infty} d\tilde{s} \,
\cos\Big[\frac{\tilde{s}\sigma\Omega(\gamma_A
-\gamma_B)}{\gamma_A^2+\gamma_B^2}\Big]\nonumber\\ &
~~~\times\exp\Big[\frac{-\tilde{s}^2}{2(\gamma_A^2+\gamma_B^2)}\Big]\Big[(\Delta{d})^2+R_A^2+R_B^2-2R_AR_B\cos\big(\tilde{s}\omega\sigma\big)-\sigma^2(\tilde{s}+i\epsilon)^2\Big]^{-1}\;.
\end{align}

Due to the complexity of the integrand in Eq.~(\ref{Xint1-1}) and
Eq.~(\ref{Xint1-2}), it is hard to obtain  analytical results.
Therefore,  numerical evaluations are needed. Nevertheless, it is
still quite a challenge to obtain  numerical results since the
integrand  is a oscillatory function with singularities.
Fortunately, the Wightman functions in fact are well-defined
distributions~\cite{EDU:2016-1,Bogolubov:1990}. So,  some techniques
of a distribution function integral (in the Cauchy principal sense,
see Appendix~(\ref{Derivation-PD})) can be utilized to obtain the
correct results, and some special numerical integration methods or
strategies (e.g., composite Simpson's rule and Legendre-Gauss
quadrature) can be of help as well. Once the values of
transition probabilities and $X^{\parallel}$ are evaluated correctly, then
the concurrence can be straightforwardly  obtained from
Eq.~(\ref{con1}).

 For simplicity, we first consider the impact of
acceleration on entanglement harvesting in the situation where two
detectors are rotating with the same  acceleration and trajectory radius, i.e,
$a_A=a_B=a\;,R_A=R_B=R\;$ (or equivalently,
$v_A=v_B\;,|\omega_A|=|\omega_B|$). For such a situation, it is easy
to judge that $\gamma_A=\gamma_B=\gamma$. Then  the non-local correlation Eq.~(\ref{Xint1-1}) can be written as
\begin{align}\label{Xint1-3}
X^{\parallel}\big|_{a,R}&=- \frac{\lambda^2\sigma^2}{2\pi^2\gamma^2}
\int_{-\infty}^{\infty} d\tilde{u} \int_{0}^{\infty} d\tilde{s} \,
\exp\big[(2\tilde{s}\tilde{u}-\tilde{s}^2-2\tilde{u}^2)/(2\gamma^{2})\big]
\exp\big[i\Omega\sigma (\tilde{s}-2\tilde{u})/\gamma\big] \nonumber\\
& \times\Big\{(\Delta{d})^2+4R^2\sin^2\big[(\tilde{u}
\omega_A-\tilde{u}\omega_B+\tilde{s}\omega_B)\sigma/2\big]-\sigma^2(\tilde{s}+i\epsilon)^2\Big\}^{-1}\;.
\end{align}
 If the two
detectors are completely comoving ($\omega_A=\omega_B=\omega$), then
the double integral Eq.~(\ref{Xint1-3}) can be simplified further to a
one-dimensional integral  by integrating $\tilde{u}$  first
\begin{equation}\label{Xint1-4}
X^{\parallel}\big|_{a,R}=- \frac{\lambda^2
\sigma^2e^{-\sigma^2\Omega^2}}{2\pi^{3/2}\gamma} \int_{0}^{\infty}
d\tilde{s}
\frac{e^{-\tilde{s}^2/(4\gamma^2)}}{\Delta{d}^2+4R^2\sin^2(\tilde{s}\omega\sigma/2)-\sigma^2(\tilde{s}+i\epsilon)^2}\;.
\end{equation}

%%%%%%%%%%%%%%%%%%%
\begin{figure*}[!ht]
\centering
\subfigure[]{\label{convsdall11}
\includegraphics[width=0.46\textwidth]{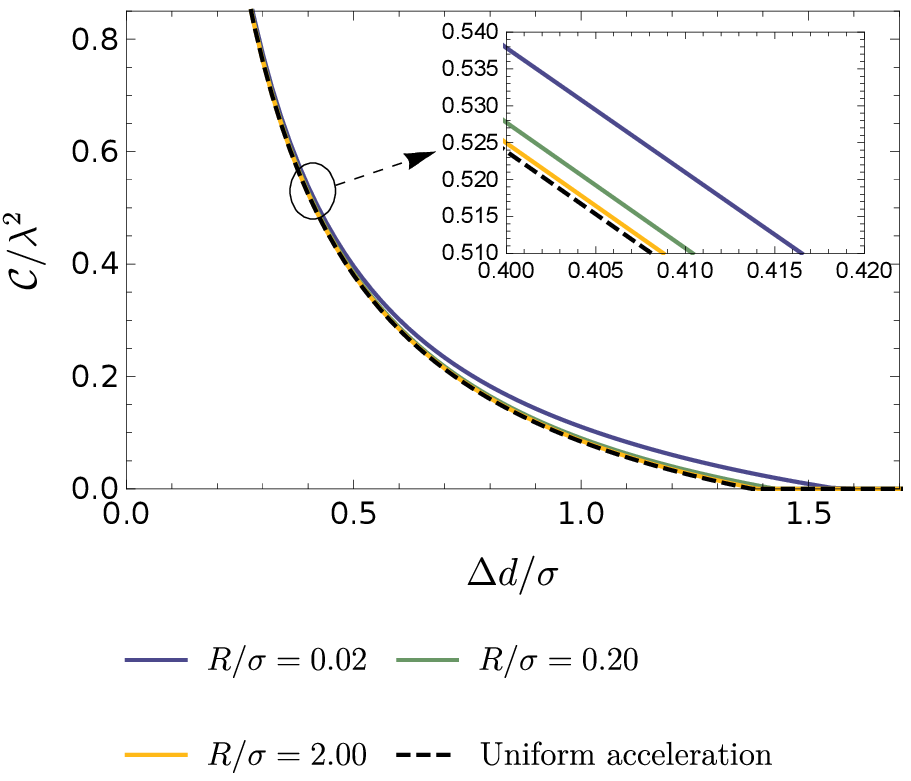}}\hspace{0.03\textwidth}\subfigure[]{\label{convsdall12}
\includegraphics[width=0.46\textwidth]{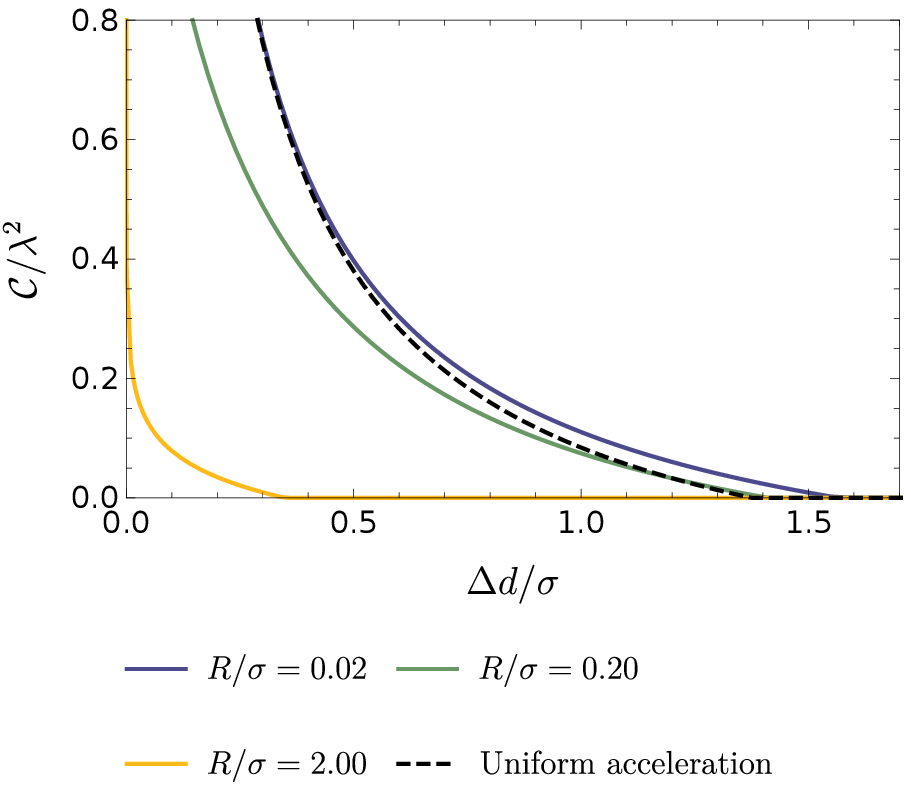}}
\caption{The  concurrence ${\cal{C}}(\rho_{AB})/\lambda^2$ is plotted as a function of
$\Delta{d}/\sigma$. Suppose two such identical
circularly accelerated detectors are coaxially rotating in an
equivalent acceleration and radius, i.e., $a_A=a_B=a$ and
$R_A=R_B=R$, setting $a\sigma=1.0$ and $\Omega\sigma=0.10$. Note
that as for (a) two detectors have a same angular velocity
($\omega_A=\omega_B$)  and for (b) they have mutually opposite
angular velocities ($\omega_A=-\omega_B$). Here, the additional
black dashed line describes the linearly uniformly accelerated
situation.}\label{convsdall1}
\end{figure*}

%%%%%%%%%%%%%%%%%%%%%%%%%%%%%%%%
To facilitate a comparison with the situation of linear uniformly
accelerated motion,  we  consider the following trajectory for
uniform acceleration
\begin{align}\label{traj-ua}
&x_A:=\{t=a^{-1}\sinh(a\tau_A)\;,~x=a^{-1}\cosh(a\tau_A)\;,~y=0\;,~z=0\}\;,
\nonumber\\
&x_B:=\{t=a^{-1}\sinh(a\tau_B)\;,~x=a^{-1}\cosh(a\tau_B)\;,~y=0\;,~z=\Delta{d}\}\;,
\end{align}
where the symbol $a$ still denotes the magnitude of uniform
acceleration and $\Delta{d}$  stands for the separation between
two detectors. Similarly, the transition probabilities and  the
non-local correlations represented by $X$ can
also be straightforwardly calculated by substituting the trajectories
into Eq.~(\ref{probty}) and Eq.~(\ref{defX}) (the general expression
for that uniform acceleration has been studied in
Ref.~\cite{Salton:2015}).

 In Fig.~(\ref{convsdall1}), the concurrence is plotted as a function of the separation $\Delta{d}$ in the unit of $\sigma$.
 As we will see that the entanglement (concurrence) in general is a fast decaying
 function of $\Delta{d}/\sigma$ irrespective of the direction of angular
 velocity. It means that a large separation   generally inhibits the detectors from harvesting entanglement.
 For $\omega_A=\omega_B$, the entanglement harvested by  circularly accelerated detectors
 is not expectedly sensitive to the trajectory radius, decaying  a little more slowly  than that for uniformly accelerated
 situation with  increasing separation $\Delta{d}$.  However, for $\omega_A=-\omega_B$,  the decaying behavior of entanglement will  be more sensitive to the radius. Especially for a large radius ($R_D/\sigma>1$), it  will rapidly fall  to zero with  increasing $\Delta{d}/\sigma$ since the size of their circular trajectories can  enlarge the average separation between two detectors.
%%%%%%%%%%%%%%%%%%
\begin{figure*}[!ht]
\centering
\subfigure[]{\label{convsARall11}
\includegraphics[width=0.45\textwidth]{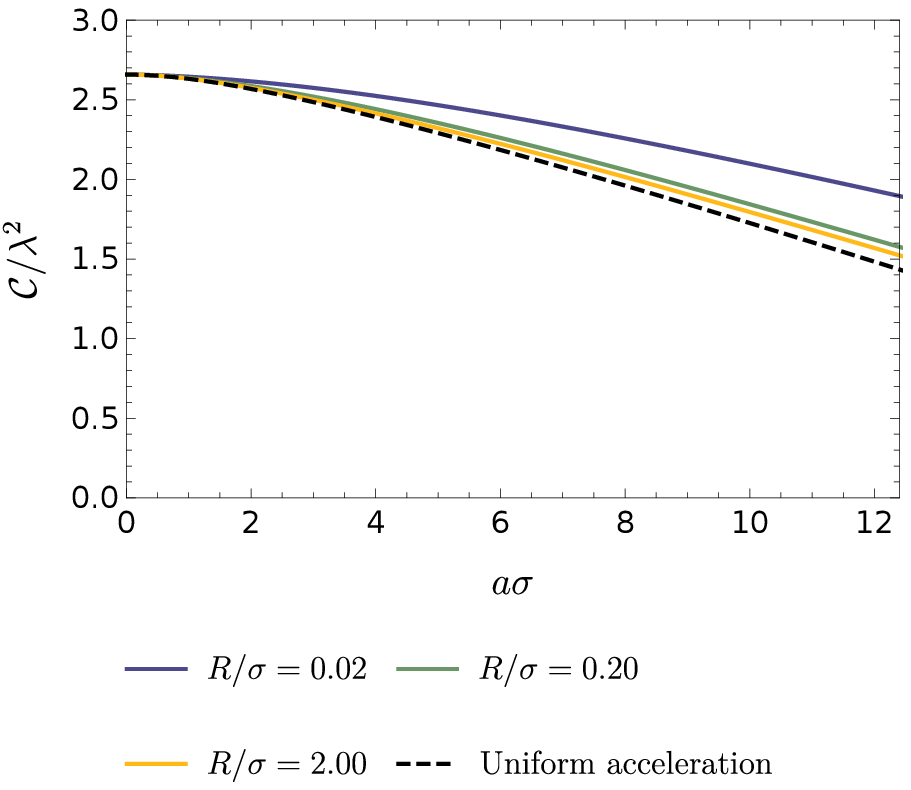}}\hspace{0.05\textwidth}\subfigure[]{\label{convsARall12}
\includegraphics[width=0.45\textwidth]{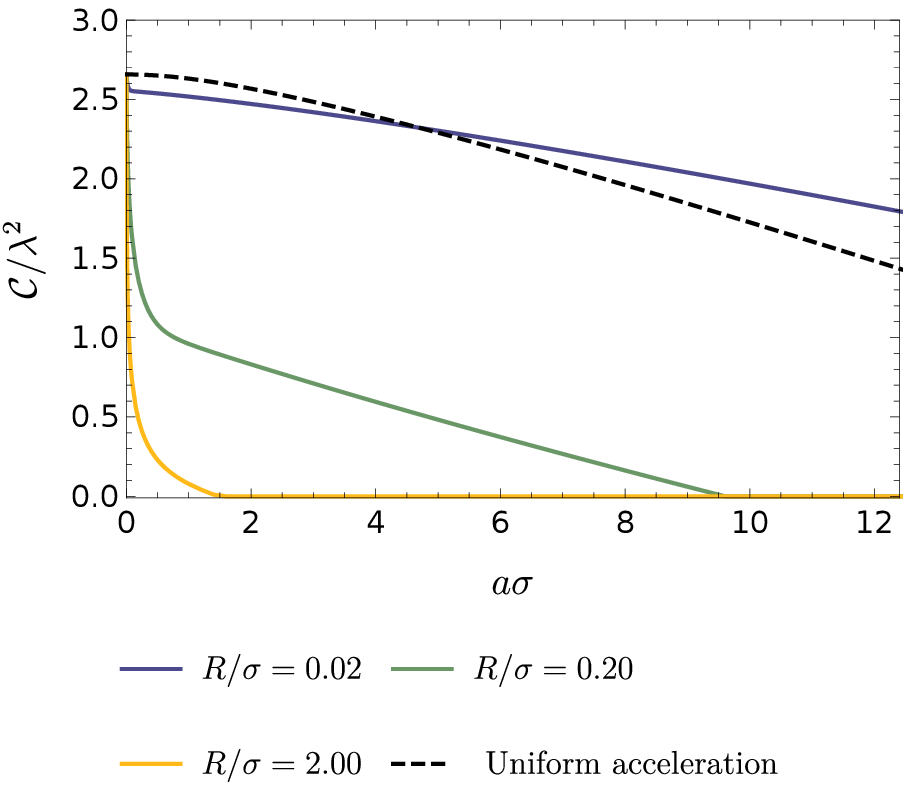}}
\caption{Assuming $a_A=a_B=a$ and $R_A=R_B=R$ in coaxial rotating
motion, the concurrence ${\cal{C}}(\rho_{AB})/\lambda^2$ is plotted
as a function of $a\sigma$, satisfying $\Delta{d}/\sigma=0.10$ and
$\Omega\sigma=0.10$. Here we have set $\omega_A=\omega_B$ in (a) and
$\omega_A=-\omega_B$ in (b). The additional black dashed lines in
both plots are identical, which describe the uniformly accelerated
situation for comparison.}\label{convsARall1}
\end{figure*}

\begin{figure}[!ht]
\centering{
\includegraphics[width=0.8\textwidth]{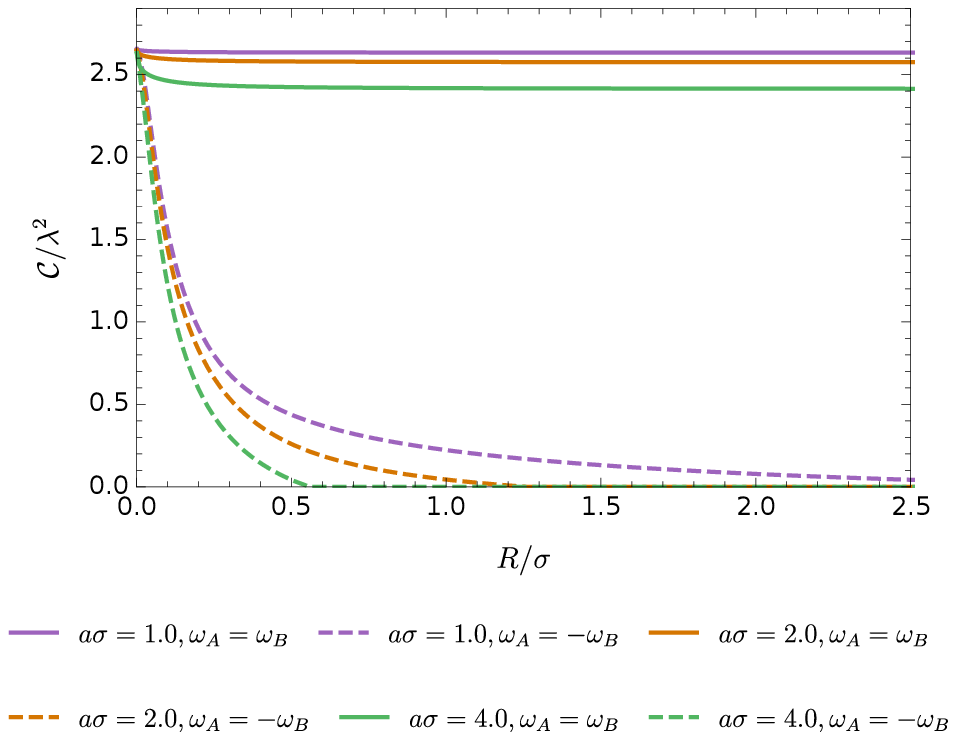}}
\caption{The concurrence ${\cal{C}}(\rho_{AB})/\lambda^2$ is plotted
as a function of the radius in coaxial rotation with $a_A=a_B=a$,
$R_A=R_B=R$, $\Omega\sigma=0.10$ and
$\Delta{d}/\sigma=0.10$.}\label{convsRall1}
\end{figure}

In Fig.~(\ref{convsARall1}),  the dependence  of
 ${\cal{C}}(\rho_{AB})/\lambda^2$  on the acceleration with various circular trajectory radii has been displayed. If such two
detectors is in co-rotation  with $\omega_A=\omega_B$, it is easy to find that the
larger the trajectory radius $R/\sigma$  is, the faster the entanglement
decays with the increasing acceleration $a\sigma$. While if two detectors are in counter-rotation  with an equal angular velocity
($\omega_A=-\omega_B$), a large radius renders the entanglement sharply decay with  increasing acceleration, while a very small $R/\sigma$ suppresses the decay as compared to the large radius case as  $a\sigma$ increases, even making the entanglement decay much more slowly than that for the situation of uniform acceleration.

To understand the above referred characters, we recall that  the
concurrence is determined by the competition between  the
non-local correlation represented by $X$ and
transition probabilities, which means that decreasing $X$ or
increasing the transition probability may render the entanglement
decrease.  According to Eq.~(\ref{Xint1-1}), the non-local correlation  $X^{\parallel}$  is mainly dependent
upon the average separation  between two detectors, while  the transition probability is  dominantly  determined by the value of acceleration as shown in  Fig.~(\ref{PAPBvsA}). For the case of co-rotation ($\omega_A=\omega_B$),  the separation
is always fixed, and the nearly uncharged value of $X^{\parallel}$ and the increase of transition probability will make the concurrence monotonically
decay over  the entire range of $a\sigma$. However, according to Fig.~(\ref{PAPBvsA}) and Fig.~(\ref{PAPBvsTed}), we can see that a large trajectory radius would make the transition probability greater than that for a small radius, i.e., for not too small acceleration the detector rotating along with a larger circular trajectory may observe stronger thermal-like noise which can hinder it from harvesting entanglement. Thus, the larger the trajectory radius is, the faster the entanglement decays with the increasing acceleration.  As for the case of counter-rotation ($\omega_A=-\omega_B$), if the trajectory radius is comparable with $\Delta{d}$, the average separation between two detectors may increase during the finite duration time, which would make the non-local correlation $X^{\parallel}$ decrease sharply.  Then the sharply decreased $X^{\parallel}$ and  thermal-like noises  render the  harvested entanglement rapidly decay to zero as the acceleration increases. However, for a vanishingly small radius ($R\ll\Delta{d}$) in the counter-rotation situation, the decreased amount of $X^{\parallel}$  is tiny due to the slight change of the separation between two detectors, hence the concurrence, analogous to the situation of co-rotation, will be  governed by the value of the transition probability.  Since the uniformly accelerated detectors observe stronger thermal-like noise for not too small acceleration (see Fig.~(\ref{PAPBvsA}) and Fig.~(\ref{PAPBvsTed})), then the entanglement harvested by circularly accelerated detectors would decay much more slowly than that by uniformly accelerated detectors. We also plot how the entanglement depends on the trajectory radius in Fig.~(\ref{convsRall1}) as a supplement. One may find that the trajectory radius would play an important inhibiting role in entanglement harvesting  in the counter-rotation situation.

%%%%%%%%%%%%%%%%%%%%%%%%%%
 Now, let us turn to the question  as to what
happens to entanglement harvesting in the situation where the two
detectors  are coaxially rotating with different values of
acceleration or angular velocity. For simplicity,  we suppose that such
two detectors are in concentric circular motion in $xy$-plane, i.e.,
$\Delta{d}=0$, along the trajectory~(\ref{traj2}). According to
Eq.~(\ref{Xint1-1}),  the
corresponding concurrence can be straightforwardly obtained  via
numerical evaluations.
%%%%%%%%%%%%%%
\begin{figure}[!ht]
\centering{
\includegraphics[width=0.85\textwidth]{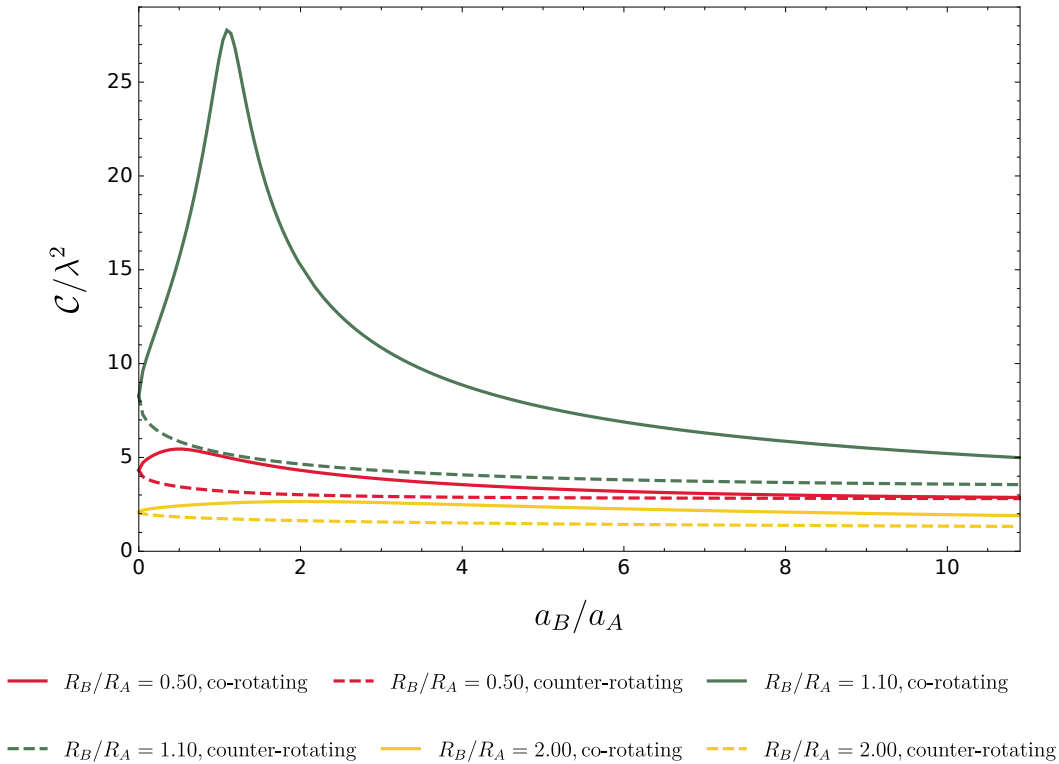}}
\caption{The concurrence ${\cal{C}}(\rho_{AB})/\lambda^2$ is plotted
as a function of $a_B/a_A$ for two detectors in concentric circular
motion in $xy$ plane. Here, we have set $R_A/\sigma=0.10$, $\omega_A\sigma=1.00$,
$\Omega\sigma=0.10$ and $\Delta{d}=0$ . The dashed lines correspond to the situation of two detectors rotating in opposite directions.}\label{convsAbAAall1}
\end{figure}
\begin{figure}[!ht]
\centering{
\includegraphics[width=0.8\textwidth]{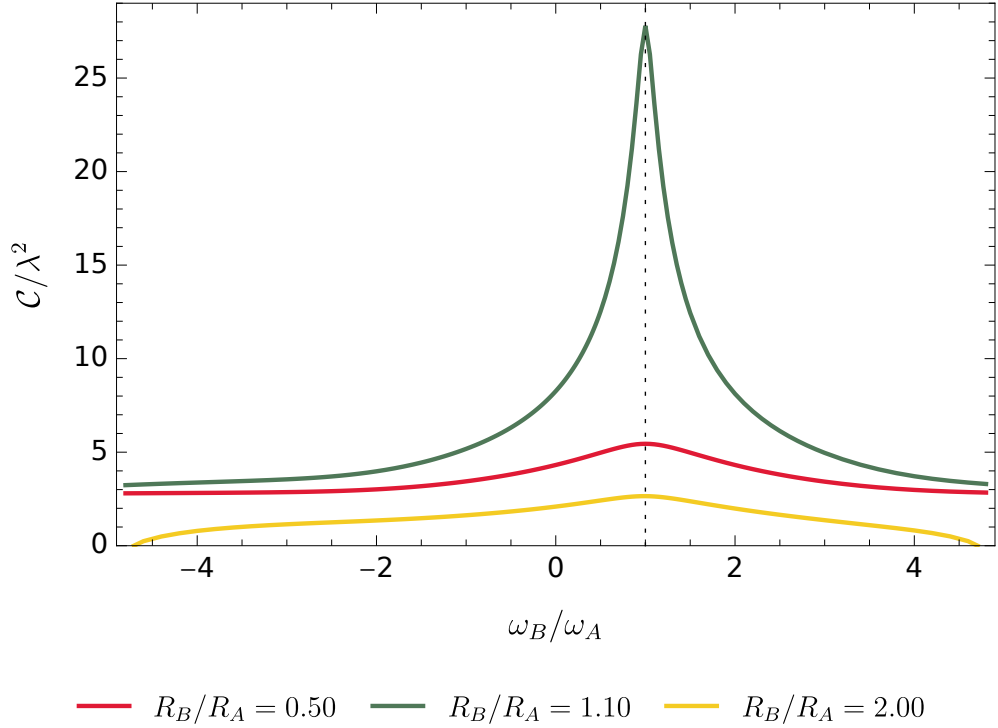}}
\caption{The concurrence ${\cal{C}}(\rho_{AB})/\lambda^2$ is plotted
as a function of $\omega_B/\omega_A$ for  two detectors
concentrically rotating in $xy$ plane. Here we have set
$\Delta{d}=0$, $\omega_A\sigma=1.00$, $R_A/\sigma=0.10$ and
$\Omega\sigma=0.10$. The vertical dotted line denotes the point
$\omega_B/\omega_A=1$ at which ${\cal{C}}(\rho_{AB})/\lambda^2$ will
take the maximum value. }\label{convsdwall1}
\end{figure}
%%%%%%%%%%%%%%%%%%%%%%%%

As shown in Fig.(\ref{convsAbAAall1}), it is  difficult to extract the entanglement from the vacuum state of the quantum field throughout a large region of $a_B/a_A$, irrespective of the direction of angular velocity. Most intriguingly, we find that ${\cal{C}}(\rho_{AB})/\lambda^2$ is not zero but taking a finite
value in the limit of $a_B=0$ (or equivalently, $\omega_B=0$). Such nonzero result at this special point tells us that the entanglement is likely to be still harvested by  detectors $A$ and $B$ which are respectively at rest and in circular motion. In addition, it should be pointed out the harvested entanglement in the counter-rotation situation can approach its maximum at $a_B=0$. Nevertheless, we expect that the peak value of entanglement for a fixed radius ratio $R_A/R_B$ should be  achieved at  the point of $\omega_B=\omega_A$ in the co-rotation situation.

In order to check this,  and gain a better understanding of how the entanglement
depends on the difference between the angular velocities of two
detectors, we plot the concurrence as a function
$\omega_B/\omega_A$ in Fig.~(\ref{convsdwall1}). As we have seen
that the smaller the radius difference between two detectors, the more the entanglement harvested. Remarkably, when two detectors
are synchronously rotating ($\omega_B=\omega_A$), the concurrence
would certainly take the maximum value since the separation between two
detectors always remains a minimum value, which is consistent with
our intuitive perception that  synchronously concentrically rotating
detectors will extract the most entanglement from the vacuum state of a
quantum field.

\subsection{ the situation of mutually perpendicular rotation axes}
Let us now consider that the two detectors are rotating in mutually
perpendicular planes. The trajectories  are given by
\begin{align}\label{traj3}
&x_A:=\{t=\tau_A\gamma_A\;, x=R_A\cos(\omega_A\tau_A\gamma_A)\;,~
y=R_A\sin(\omega_A\tau_A\gamma_A)\;,z=0\}\;,
\nonumber\\
&x_B:=\{t=\tau_B\gamma_B\;,
x=R_B\cos(\omega_B\tau_B\gamma_B)+\Delta{d}\;,
~y=0\;,z=R_B\sin(\omega_B\tau_B\gamma_B)\}.
\end{align}
Note that the transition probabilities can still be obtained by
carrying out Eq.~(\ref{PAPB}) in numerical evaluation. Here, we use $X^{\perp}$ to denote  $X$  in the case where the
trajectories~(\ref{traj3}) of two detectors are in  mutually perpendicular orbital
planes.
 Similarly,
according to Eq.~(\ref{defX}), the parameter $X^{\perp}$ can be written as
\begin{align}\label{defX-v}
X^{\perp}&=-
\frac{\lambda^2\sigma^2}{4\pi^2\gamma_A\gamma_B}
\int_{-\infty}^{\infty}  d\tilde{u} \int_{0}^{\infty} d\tilde{s} \,
\bigg\{\exp\Big[\frac{-\gamma_A^2\tilde{u}^2-\gamma_B^2(\tilde{s}-\tilde{u})^2}{2\gamma_A^2\gamma_B^2}\Big]
\exp\Big[\frac{i(\tilde{s}-\tilde{u})\sigma\Omega}{\gamma_A}-\frac{i\tilde{u}\sigma\Omega}{\gamma_B}\Big]\nonumber\\
&
\times{\tilde{f}_{AB}}(\tilde{u},\tilde{s})+\exp\Big[\frac{-\gamma_B^2\tilde{u}^2-\gamma_A^2(\tilde{s}-\tilde{u})^2}{2\gamma_A^2\gamma_B^2}\Big]
\exp\Big[\frac{i(\tilde{s}-\tilde{u})\sigma\Omega}{\gamma_B}-\frac{i\tilde{u}\sigma\Omega}{\gamma_A}\Big]{\tilde{f}_{BA}}(\tilde{u},\tilde{s})
\bigg\}\;,
\end{align}
where
\begin{align}
\tilde{f}_{AB}(\tilde{u},\tilde{s})=&\Big\{R_A^2+R_B^2-2R_AR_B\cos[(\tilde{u}-\tilde{s})\omega_A\sigma]\cos(\tilde{u}\omega_B\sigma)
-2R_A\Delta{d}\cos[(\tilde{u}-\tilde{s})\omega_A\sigma]\nonumber\\&+2R_B\Delta{d}\cos(\tilde{u}\omega_B\sigma)+\Delta{d}^2-\sigma^2(\tilde{s}+i\epsilon)^2\Big\}^{-1}\;,
\end{align}
\begin{align}
\tilde{f}_{BA}(\tilde{u},\tilde{s})=&\Big\{R_A^2+R_B^2-2R_AR_B\cos[(\tilde{u}-\tilde{s})\omega_B\sigma]\cos(\tilde{u}\omega_A\sigma)
+2R_B\Delta{d}\cos[(\tilde{u}-\tilde{s})\omega_B\sigma]\nonumber\\&-2R_A\Delta{d}\cos(\tilde{u}\omega_A\sigma)+\Delta{d}^2-\sigma^2(\tilde{s}+i\epsilon)^2\Big\}^{-1}\;.
\end{align}
It is quite a challenge to further simplify the above
expression of $X^{\perp}$ into a one-dimensional integral since the
trajectories~(\ref{traj3}) do not represent  the comoving circular
motion around a common rotating axis. However, it is easy to find
that Eq.~(\ref{defX-v}) is independent of the direction  of angular
velocity in comparison with Eq.~(\ref{Xint1-1}).
 Thus,  we can focus on the positive angular velocity for the trajectories Eq.~(\ref{traj3}).
 %%%%%%%%%%%%%%%%%%%
 \begin{figure}[!ht]
\centering{
\includegraphics[width=0.8\textwidth]{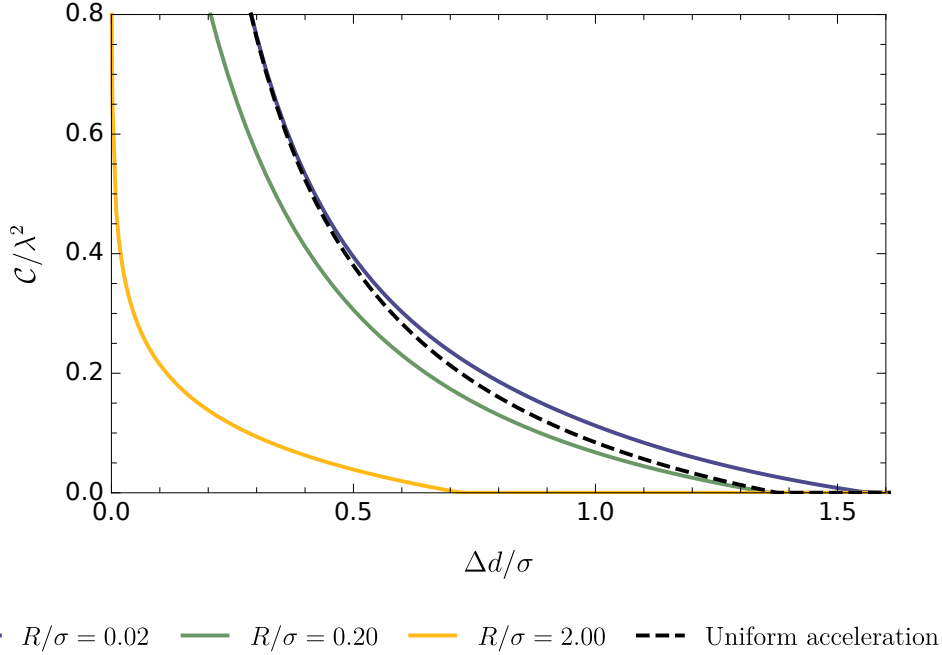}}
\caption{The concurrence  is plotted as a function of
$\Delta{d}/\sigma$ for the case of two detectors circularly rotating
with mutually perpendicular rotation axes. Here, we have set
$a_A=a_B=a$ and $R_A=R_B=R$, yielding $a\sigma=1.0$ and
$\Omega\sigma=0.10$.  The additional black dashed line describes the
uniformly accelerated situation. }\label{convsdall2}
\end{figure}

 \begin{figure}[!ht]
\centering{
\includegraphics[width=0.8\textwidth]{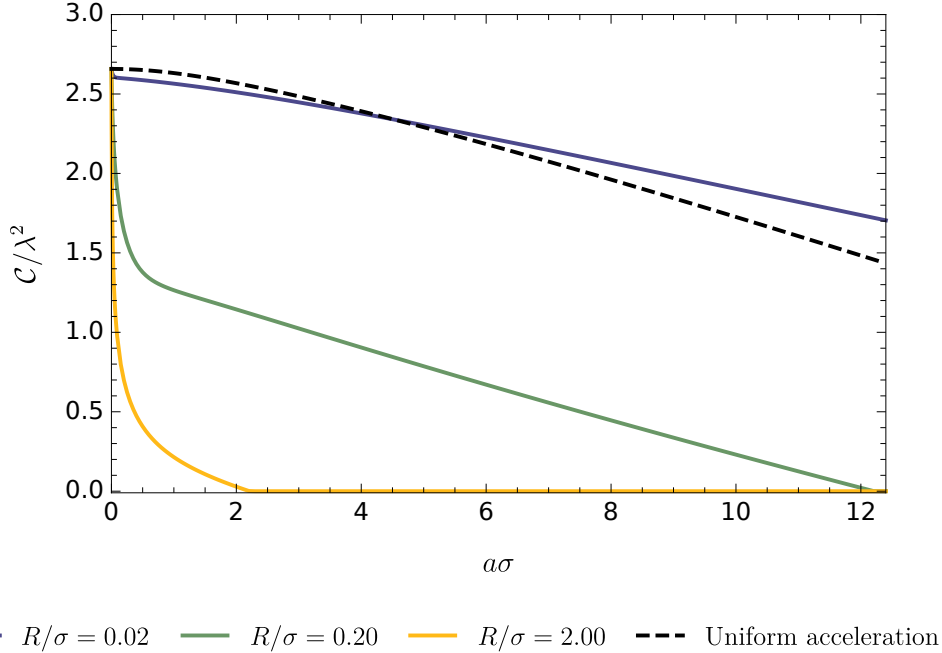}}
\caption{ As for the situation of  mutually perpendicular rotation
axes with $a_A=a_B=a$ and $R_A=R_B=R$, the concurrence
${\cal{C}}(\rho_{AB})/\lambda^2$ is plotted as a function of
$a\sigma$. The additional black dashed line describes the  situation of uniform
acceleration. Here we have set $\Delta{d}/\sigma=0.10$ and
$\Omega\sigma=0.10$. }\label{convsARall2}
\end{figure}
\begin{figure}[!ht]
\centering{
\includegraphics[width=0.8\textwidth]{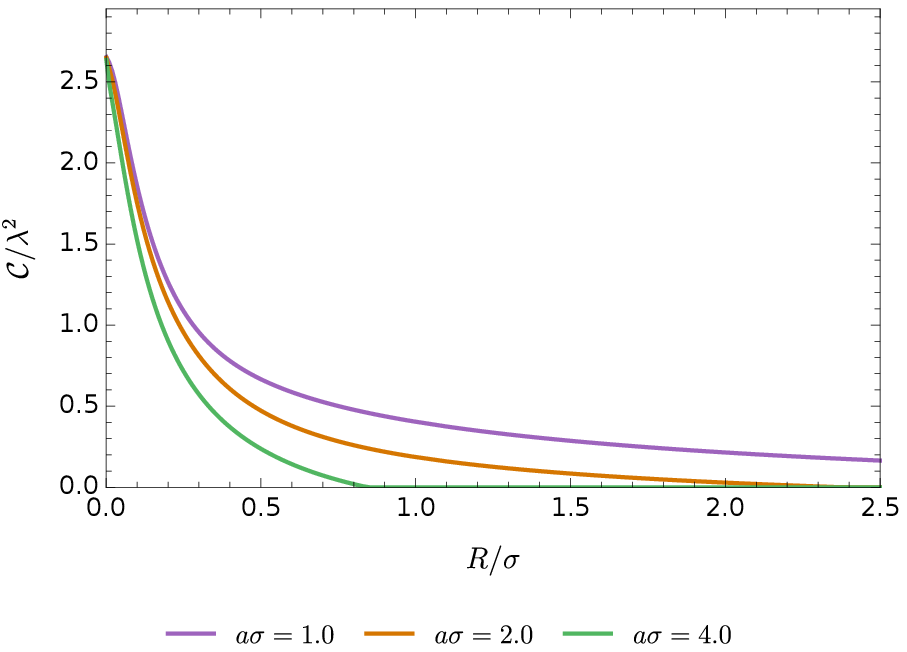}}
\caption{The concurrence vs the circular trajectory radius for  two
detectors rotating around  mutually perpendicular rotation axes.
Here, we have set $a_A=a_B=a$, $R_A=R_B=R$, $\Omega\sigma=0.10$ and
$\Delta{d}/\sigma=0.10$.}\label{convsRall2}
\end{figure}
 %%%%%%%%%%%%%%%%%%%%
For the case where two circularly accelerated detectors have  equivalent
acceleration and  trajectory radius, how the entanglement depends on the
parameters of circular motion is illustrated in
Figs.~(\ref{convsdall2})-(\ref{convsRall2}). Obviously, one can
observe an analogous harvesting behavior that looks like  the afore-studied
situation of the coaxial rotation with equal and opposite angular
velocities, though the quantitative details are different. For example, for a large trajectory radius, the harvested entanglement
for perpendicular rotations falls to zero a little more slowly than that for coaxial rotations with opposite angular velocities (see the
curves of $R/\sigma=2.00$ in Fig.~(\ref{convsdall12}) and Fig.~(\ref{convsdall2})). Similar conclusions can be obtained directly  by comparing Fig.~(\ref{convsARall12}) with  Fig.~(\ref{convsARall2}) or Fig.~(\ref{convsRall1}) with  Fig.~(\ref{convsRall2}).

%%%%%%%%%%%%%%%%
\begin{figure}[!ht]
\centering{
\includegraphics[width=0.8\textwidth]{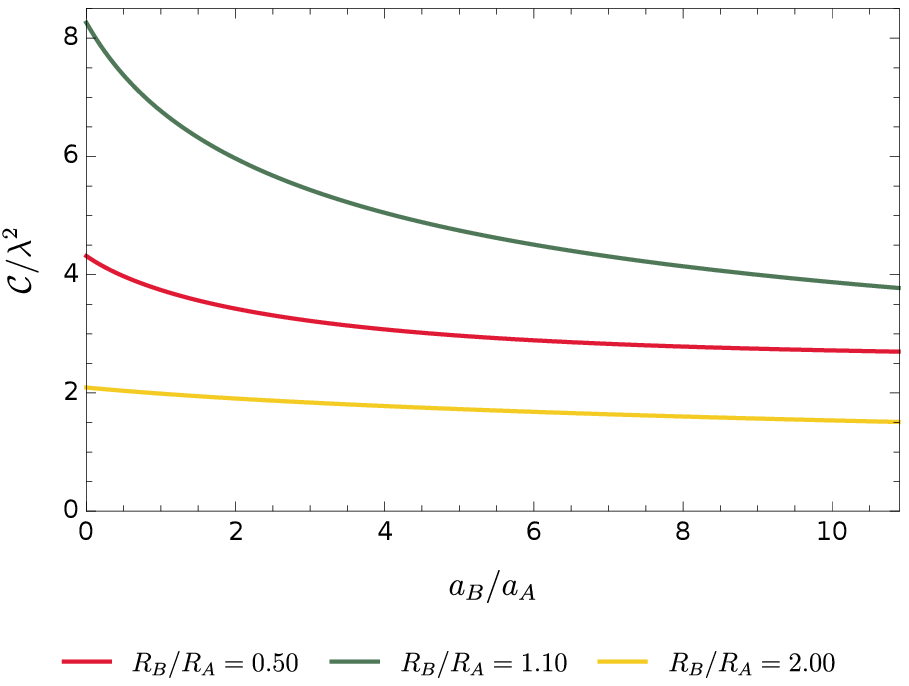}}
\caption{The concurrence ${\cal{C}}(\rho_{AB})/\lambda^2$ is plotted
as a function of $a_B/a_A$ for two detectors $A$ and $B$ circularly
rotating in $xy$ and $xz$ planes, respectively. Here, we have set
$R_A/\sigma=0.10$, $\omega_A\sigma=1.00$, $\Omega\sigma=0.10$ and $\Delta{d}=0$.} \label{convsAbAAall2}
\end{figure}
\begin{figure}[!ht]
\centering{
\includegraphics[width=0.8\textwidth]{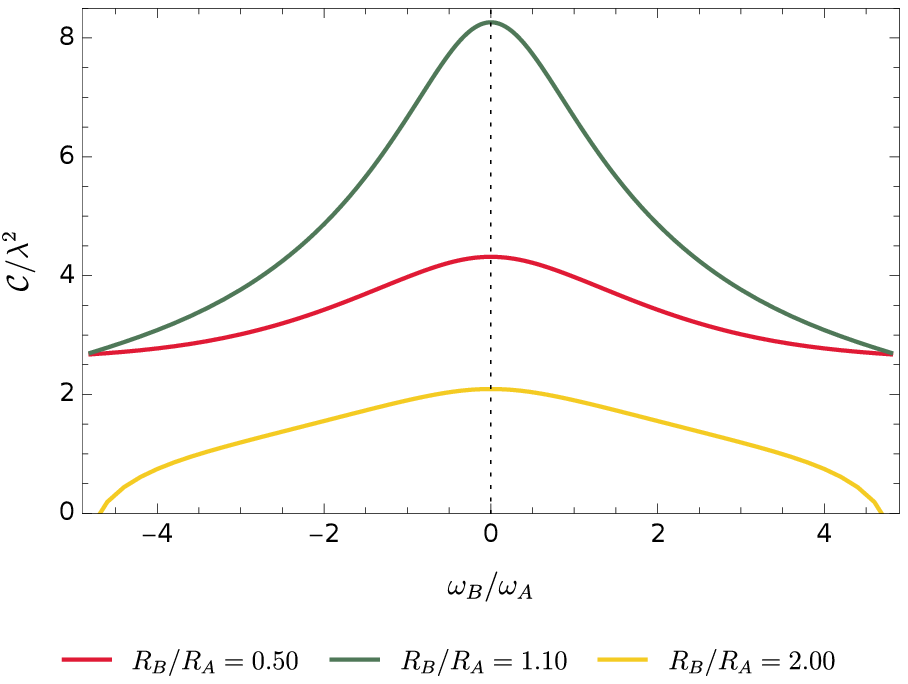}}
\caption{The concurrence ${\cal{C}}(\rho_{AB})/\lambda^2$ is plotted
as a function of $\omega_B/\omega_A$ for the detectors $A$ and $B$
concentrically rotating in $xy$ and $xz$ planes respectively. Here
we have set $\Delta{d}=0$, $\omega_A\sigma=1.00$, $R_A/\sigma=0.10$
and $\Omega\sigma=0.10$. The vertical dotted line denotes the point
$\omega_B=0$ at which ${\cal{C}}(\rho_{AB})/\lambda^2$ will take the
maximum value. }\label{convsdwall2}
\end{figure}
%%%%%%%%%%%%%%%%%%%%
In Figs.~(\ref{convsAbAAall2})-(\ref{convsdwall2}), we have considered
the situation of two detectors rotating at different accelerations
or angular velocities along the trajectories~(\ref{traj3}) with
$\Delta{d}=0$. When comparing Fig.~(\ref{convsAbAAall2}) with
Fig.~(\ref{convsAbAAall1})(the counter-rotation situation),  similar results can be obtained: the
larger  the acceleration ratio between two detectors, the
less the harvested entanglement.  In the limit of
$a_B=0$ (or $\omega_B=0$), ${\cal{C}}(\rho_{AB})/\lambda^2$ will not be vanishing but take a non-zero maximum value.  The most notable difference between Figs.(\ref{convsdwall2}) and (\ref{convsdwall1}) is that the peak of the
extracted entanglement for non-coaxial rotations is localized at
point $\omega_B/\omega_A=0$ rather than $\omega_B/\omega_A=1$ for
coaxial rotations.
\section{conclusions}
In this paper, we have performed  detailed discussions on the
properties of  the transition probability of a circularly accelerated
UDW detector coupled with the massless scalar fields with a Gaussian
switching function, and investigated the entanglement harvesting
phenomenon of two such detectors through the corresponding
harvesting protocol. With the help of numerical evaluation, we have
analyzed the influence of motion parameters on transition
 probabilities from various aspects. It was found that the transition probability of the  circularly accelerated
 detector with a larger trajectory radius is more sensitive to the magnitude of acceleration, so is the effective temperature $T_{\rm{EDR}}$.
 By a cross-comparison of the situations of circularly  and  uniformly accelerated motion,
we obtain that the transition probability and effective temperature
$T_{\rm{EDR}}$ for circularly accelerated detectors with an
extremely large radius  and that for uniformly accelerated
 detectors would behave analogously,  but for a vanishingly small linear speed and a large acceleration with not-extremely small energy gap,
 $T_{\rm{EDR}}\approx{a}_Dv_D\sqrt{1-v_D^2}/6$
 in a finite duration, which  differs  from $T_{\rm{EDR}}\approx{a_D}/(2\pi)$ for uniformly
highly accelerated  detectors in a finite duration. And it seems
that there are no anti-Unruh phenomena for circularly accelerated
detectors interacting with the massless scalar fields in Minkowski
spacetime.

For the purpose of well understanding the entanglement harvesting
phenomenon, we  focus on two special circular motion
situations, i.e., coaxial  rotation and mutually perpendicular axial
rotation. When such two identical detectors are rotating along the circular trajectories with
equivalent radius and acceleration, it was found that the harvested
entanglement in both special circular motion situations
decays with increasing  acceleration  or separation between
two detectors. By a cross-comparison of the concurrence
for circularly accelerated  and uniformly accelerated detectors, we
 find that the trajectory radius and angular velocity in circular motion would have
great effects on the entanglement harvesting phenomenon. Especially
for two circularly accelerated detectors rotating in different directions, the trajectory radius would play an
important inhibiting role in  entanglement harvesting. In addition,
it is worth pointing out that the behavior of entanglement generation for
the situation of mutually perpendicular axes is analogous to that
for the coaxial situation with $\omega_A=-\omega_B$, but the
quantitative details are different slightly.

Finally, we  have also investigated  the entanglement harvesting for
two  identical detectors circularly rotating at different
accelerations or angular velocities. The numerical results tell us
that the entanglement harvesting still occurs  in the quantum system
composed of  rest  and  circularly accelerated detectors.
Particulary, in the situation of mutually perpendicular rotating
axes (satisfying the circular orbit Eq.~(\ref{traj3})), the
extracted entanglement  will take the peak value when one detector
is at rest and the other is circularly accelerated.  Thus,  an
interesting question arises as to what the entanglement harvesting
exactly behaves when two detectors are in completely different
motion status. For example, one detector keeps at rest while the
other is uniformly accelerated. Such a situation  is analogous to
that  of one detector falling into black hole while the other stays
outside. So, a further in-depth study on the entanglement harvesting
for two detectors which are separated by a horizon in the Rindler
spacetime or some curved spacetimes with a black hole is
particularly desirable, which we would rather leave to a future
work.

\begin{acknowledgments}
 This work was supported by the National Natural Science Foundation of China under Grant  No.11690034.
\end{acknowledgments}
\appendix
\section{Derivation of $P_D$ and $X$}\label{Derivation-PD}
In this appendix we will derive  the explicit form of the transition
probability $P_D$ and $X$ from Eq.~(\ref{probty}) and
Eq.~(\ref{defX}) respectively.
\subsection{The transition probability $P_D$}\label{Derivation-PD1}
Recalling Eq.~(\ref{probty}), and letting $u=\tau_D$ and
$s=\tau_D-\tau_D'$, then the transition probability  can be
rewritten as
\begin{align}\label{PA-A}
P_D&=\lambda^2\int_{-\infty}^{\infty}{du}\chi_D(u)\int_{-\infty}^{\infty}{ds}\chi_D(u-s)e^{-i\Omega{s}}W(s)\nonumber\\
&=\lambda^2\sqrt{\pi}\sigma\int_{-\infty}^{\infty}{ds}e^{-i\Omega{s}}e^{-s^2/(4\sigma^2)}W(s)\;.
\end{align}
Inserting the Wightman function Eq.~(\ref{wightman2}) into
Eq.~(\ref{PA-A}) and assuming $x=\gamma_D|\omega_D|s/2$
 lead to
\begin{align}\label{PA2-A}
P_D&=\frac{\lambda^2\sigma|\omega_D|}{8\pi^{3/2}\gamma_D}\int_{-\infty}^{\infty}dx\frac{{e}^{-i
2x\Omega/(\gamma_D|\omega_D|)}e^{-x^2/(\gamma_D^2\sigma^2\omega_D^2)}}{R_D^2\omega_D^2\sin^2x-(x-i\epsilon)^2}\;,
\nonumber\\
&=\frac{\lambda^2\sigma|\omega_D|}{8\pi^{3/2}\gamma_D}\int_{-\infty}^{\infty}dx\bigg[\frac{{e}^{-i x\beta}e^{-x^2\alpha}}{v_D^2\sin^2x-(x-i\epsilon)^2}+\frac{{e}^{-i x\beta}e^{-x^2\alpha}}{(1-v_D^2)(x-i\epsilon)^2}-\frac{{e}^{-i x\beta}e^{-x^2\alpha}}{(1-v_D^2)(x-i\epsilon)^2}\bigg]\;,
\nonumber\\
&=K_D\int_{0}^{\infty}dx\frac{\cos(
x\beta)e^{-x^2\alpha}(x^2-\sin^2x)}{x^2(x^2-v_D^2\sin^2x)}-\frac{\lambda^2\sigma|\omega_D|}{8\pi^{3/2}\gamma_D}\int_{-\infty}^{\infty}dx\frac{{e}^{-i
x\beta}e^{-x^2\alpha}}{(1-v_D^2)(x-i\epsilon)^2}
\end{align}
with $\alpha:=1/(\sigma^2\omega_D^2\gamma_D^2),$
$\beta:={2\Omega}/(\gamma_D|\omega_D|)$ and
$K_D:={\lambda^2v_D^2\gamma_D|\omega_D|\sigma}/(4\pi^{3/2})$. It is
worth pointing out the integrand in the first term of
Eq.~(\ref{PA2-A}) is a regular function, namely  $i\epsilon$  can
be suppressed, while the second term can be calculated  by using the
technique of  distribution functions. Let us pause to review some
properties  for distribution functions.
 Recall that the action of a distribution $g$ on a test function $f$ is defined
 by~\cite{EDU:2016-1,Bogolubov:1990}
 \begin{equation}
\langle{g},{f}\rangle:=\int_{-\infty}^{\infty}g(x)f(x)dx\;,
\end{equation}
 which satisfies the following derivative relation
  \begin{equation}\label{dist-r}
\big\langle{\frac{dg}{dx}},{f}\big\rangle=-\big\langle{g},{\frac{df}{dx}}\big\rangle\;.
\end{equation}
For the distribution $1/x$, the action is defined as
\begin{equation}
\big\langle{\frac{1}{x}},{f(x)}\big\rangle:=PV\int_{-\infty}^{\infty}\frac{f(x)}{x}dx\;,
\end{equation}
where $PV$ denotes the integral principle value.
Then  for the distribution $1/x^2$, we can obtain~\cite{Bogolubov:1990}
\begin{equation}\label{xx-int}
\big\langle{\frac{1}{x^2}},{f(x)}\big\rangle=\big\langle{\frac{1}{x}},{\frac{df(x)}{dx}}\big\rangle
=\int_{0}^{\infty}dx\frac{f(x)+f(-x)-2f(0)}{x^2}\;.
\end{equation}
 According to the Sokhotski-Plemelj formula
 \begin{equation}
\frac{1}{x\pm{i}\epsilon}=PV\frac{1}{x}\mp{i}\pi\delta(x)\;,
 \end{equation}
 the following identity can be obtained  by differentiation
 \begin{equation}
 \frac{1}{(x\pm{i}\epsilon)^n}=\frac{1}{x^n}\pm\frac{(-1)^n}{(n-1)!}{i\pi}\delta^{(n-1)}(x)\;.
\end{equation}
 We can utilize the relation Eq.~(\ref{dist-r}) to obtain the action of both the distributions $1/x^n$  and $\delta^{(n-1)}(x)$ on a test function.
 In particular, for the distribution $\delta^{(n-1)}(x)$  one
 has~\cite{Bogolubov:1990}
 \begin{equation}\label{delta-n}
 \big\langle{\delta^{(n-1)}(x)},{f(x)}\big\rangle=(-1)^{n-1}f^{(n-1)}(0)\;.
 \end{equation}
Now, let us  return to the second integral of Eq.~(\ref{PA2-A}), we
have
\begin{align}\label{PA3-A}
&-\frac{\lambda^2\sigma|\omega_D|}{8\pi^{3/2}\gamma_D}\int_{-\infty}^{\infty}dx\frac{{e}^{-i
x\beta}e^{-x^2\alpha}}{(1-v_D^2)(x-i\epsilon)^2}
\nonumber\\
&=-\frac{\lambda^2\sigma|\omega_D|\gamma_D}{8\pi^{3/2}}\int_{-\infty}^{\infty}dx{e}^{-i
x\beta}e^{-x^2\alpha}\Big[\frac{1}{x^2}-i\pi\delta^{(1)}(x)\Big]
\nonumber\\
&=-\frac{\lambda^2\sigma|\omega_D|\gamma_D}{8\pi} \Big[\sqrt{\pi}\beta \rm{Erfc}\big(\frac{\beta}{2\sqrt{\alpha}}\big)
-2e^{-\beta^2/(4\alpha)}\sqrt{\alpha}\Big]\nonumber\\
&=\frac{\lambda^2}{4\pi}\Big[e^{-\Omega^2\sigma^2}-\sqrt{\pi}\Omega\sigma
\,\rm{Erfc}\big(\Omega\sigma\big)\Big]\;,
\end{align}
where we have used the identities Eqs.~(\ref{xx-int})
and~(\ref{delta-n}) in the last but one step. Thus, combining Eqs.
(\ref{PA2-A}) and~(\ref{PA3-A}) yields the transition
probability given in Eq.~(\ref{PAPB}).

Armed with the proposed technique of  distribution functions, we can
similarly get the expression of the  transition probability  for a
 uniformly accelerated detector with the
trajectory~(\ref{un-traj}), and the result is
\begin{align}\label{PApB-UA}
P_D^{\rm{UA}}=\frac{\lambda^2a_D\sigma}{4\pi^{3/2}}\int_{0}^{\infty}dx
\frac{(\sinh^2x-x^2)\cos(2x\Omega/{a_D})}{x^2\sinh^2x}{e}^{-\frac{x^2}{a_D^2\sigma^2}}+\frac{\lambda^2}{4\pi}\Big[e^{-\Omega^2\sigma^2}-\sqrt{\pi}\Omega\sigma
\,\rm{Erfc}\big(\Omega\sigma\big)\Big]\;,
\end{align}
where the superscript ``$\rm{UA}$"  means the uniform acceleration
situation.
\subsection{The expression of $X$}\label{Derivation-PD2}
We begin from the definition of $X$ in Eq.~(\ref{defX}), which can
be rewritten as
\begin{align}
X=&-\frac{\lambda^2}{\gamma_A\gamma_B}
\int_{-\infty}^{\infty}dt\int_{-\infty}^{t}dt'
\bigg[\chi_B(\tau_B(t)) \chi_A(\tau_A(t'))
 e^{-i(\Omega {t}/\gamma_B+\Omega{t'}/\gamma_A)} W\!\left(x_A(t'), x_B(t)\right) \nonumber \\
 & \quad + \chi_A(\tau_A(t)) \chi_B(\tau_B(t'))  e^{-i( \Omega{t}/\gamma_A  +
  \Omega{t'}/\gamma_B)} W\!\left(x_B(t'),x_A(t) \right)  \bigg]\nonumber\\
 =& -\frac{\lambda^2\sigma^2}{\gamma_A\gamma_B}
\int_{-\infty}^{\infty}  d\tilde{u} \int_{0}^{\infty} d\tilde{s} \,
\bigg[e^{-\tilde{u}^2(\gamma_B^{-2}+\gamma_A^{-2})/2}e^{-\tilde{s}^2/2\gamma_A^2}
e^{\tilde{s}\tilde{u}/\gamma_A^2} e^{-i\tilde{u}\Omega\sigma[\gamma_B^{-1}+\gamma_A^{-1}]}e^{i\tilde{s}\Omega\sigma/\gamma_A}
W\!\left(x_A( t' ), x_B(t )\right)\nonumber\\
&\quad+e^{-\tilde{u}^2(\gamma_A^{-2}+\gamma_B^{-2})/2}e^{-\tilde{s}^2/2\gamma_B^2}
e^{\tilde{s}\tilde{u}/\gamma_B^2}
e^{-i\tilde{u}\Omega\sigma[\gamma_A^{-1}+\gamma_B^{-1}]}e^{i\tilde{s}\Omega\sigma/\gamma_B}
W\!\left(x_B( t' ), x_A(t )\right) \bigg]\;,\label{defX-A}
 \end{align}
 where we have assumed
 $\tilde{u}=t/\sigma,\tilde{s}=(t-t')/\sigma\;$
 in the last step.  Therefore, the expression of $X$ can be straightforwardly obtained
 by substituting the Wightman function associated with the corresponding trajectory of two detectors into Eq.~(\ref{defX-A}).
 In particular, if the Wightman function depends only on the difference
between its two arguments (i.e., the Wightman function is only
dependant on $\tilde{s}$ ), Eq.~(\ref{defX-A}) can be further
simplified into a one-dimensional integral via integrating $\tilde{s}$
first
\begin{align}
X=&-
\frac{\sqrt{2\pi}\lambda^2\sigma^2}{\sqrt{\gamma_A^2+\gamma_B^2}}
\exp\Big[\frac{-\sigma^2\Omega^2(\gamma_A+\gamma_B)^2}{2\gamma_A^2+2\gamma_B^2}\Big]\int_{0}^{\infty}
d\tilde{s} \bigg\{ \exp\Big[\frac{i\tilde{s}\sigma\Omega(\gamma_A
-\gamma_B)}{\gamma_A^2+\gamma_B^2}\Big]\exp\Big[\frac{-\tilde{s}^2}{2(\gamma_A^2+\gamma_B^2)}\Big]\nonumber\\&\times
{W}\!\left(x_A( t' ), x_B(t
)\right)+\exp\Big[\frac{i\tilde{s}\sigma\Omega(\gamma_B
-\gamma_A)}{\gamma_A^2+\gamma_B^2}\Big]\exp\Big[\frac{-\tilde{s}^2}{2(\gamma_A^2+\gamma_B^2)}\Big]W\!\left(x_B(
t' ), x_A(t )\right) \bigg\}\;.
\end{align}
Then it is straightforward to obtain  Eq.~(\ref{Xint1-2}) and Eq.~(\ref{Xint1-4})  by utilizing  the explicit expression of  the Wightman function.

\section{Some approximate results of $P_D$ and $T_{\rm{EDR}}$}\label{approx}
 To facilitate  discussions on the possible thermalization process of circularly accelerated detectors, we here derive the expressions which are needed to approximately evaluate both the transition probability  and the
EDR temperature in some special cases.
\subsection{approximate forms of $P_D$}\label{approx1}
For convenience, the parameters $\alpha$ and $\beta$ in
Eq.~(\ref{PAPB}) can be rewritten in terms of  acceleration $a_D$
and the  Lorentz factor $\gamma_D$ , as
\begin{equation}\label{PDas-alph}
\alpha=\frac{\gamma_D^2-1}{a_D^2\sigma^2}\;,~~\beta=\frac{2\Omega\sqrt{\gamma_D^2-1}}{a_D}\;.
\end{equation}
In the special case of an extremely large acceleration
($a_D\sigma\gg\gamma_D$ or $a_D\sigma\gg{R_D}/\sigma$), it follows
that $\alpha\rightarrow0$. Then  the
Gaussian switching function can be simply  dropped, therefore, it
would be better to work from  the first line of Eq.~(\ref{PA2-A})
rather than Eq.~(\ref{PAPB}), then the transition probability
takes a simple form
\begin{equation}\label{PDas-1}
P_D\approx\frac{\lambda^2\sigma|\omega_D|}{8\pi^{3/2}\gamma_D}\int_{-\infty}^{\infty}dx\frac{{e}^{-i
x\beta}}{v_D^2\sin^2x-(x-i\epsilon)^2}\;.
\end{equation}
In principle, Eq.~(\ref{PDas-1}) can be performed by utilizing the
residue theorem in the complex plane.  However, the equation
$v_D^2\sin^2x=x^2$  is not analytically solvable in the complex
plane, therefore, the poles of the integrand in Eq.~(\ref{PDas-1}) can
not be obtained exactly %which leads to no analytical integral
result. If both the speed and the energy gap are not vanishingly
small ($\beta$ is not  small too, i.e., $\beta>1$), the pure
imaginary poles of the integrand near $x=0$ would give the most
important contribution to the integral. So we can expand the sine
function to find the poles with the smallest imaginary part (such
trackable treatment in details can also be found in
Refs.\cite{Bell:1983,Audretsch :1995}), leading to
$v_D^2\sin^2x-x^2$ satisfying an approximate form
$(v_D^2-1)x^2-v_D^2x^4/3$, i.e.,
\begin{equation}\label{PDas-2}
P_D\approx\frac{\lambda^2\sigma|\omega_D|}{8\pi^{3/2}\gamma_D}\int_{-\infty}^{\infty}dx\frac{{e}^{-i
x\beta}}{(v_D^2-1)(x-i\epsilon)^2-v_D^2x^4/3}\;.
\end{equation}
Thus, we can obtain the approximate form of $P_D$ via appropriate
contour integration, and after some algebraic manipulation,
 Eq.~(\ref{PDas-2}) finally becomes
\begin{equation}\label{PDas-re}
P_D\approx\frac{a_D\sigma\lambda^2e^{-2\sqrt{3}|\Omega|/a_D}}{8\sqrt{3\pi}}+\theta(-\Omega)\frac{|\Omega|\sigma\lambda^2}{2\sqrt{\pi}}\;,
\end{equation}
where $\theta(x)$ represents the unit step function. Here, we have
used the relation
$|\omega|=a_D(1-{v_D}^2)/v_D=a_D/(\gamma_D\sqrt{\gamma_D^2-1})$.
Therefore, for an  extremely large acceleration with high speed ( $a_D\sigma\gg\gamma_D\gg1$ and $\left| \Omega \right| / a_D \ll 1$), $P_D\approx
{a_D\sigma\lambda^2}/{(8\sqrt{3\pi})}$~\cite{Doukas:2010}.

In the case of a small acceleration with high speed or extremely
large radius ( $\gamma_D\gg a_D\sigma$,
$\gamma_D\gg1\gg{a}_D|\Omega|\sigma^2$), one may find that
$\alpha\gg1$ and $\beta\gg1$ for not too small energy gap. Then the
first term in Eq.~(\ref{PAPB}) can be written as
\begin{align}\label{PDas-3}
I&=K_D\int_{0}^{\infty}dx\frac{\cos(
x\beta)e^{-x^2\alpha}(x^2-\sin^2x)}{x^2(x^2-v_D^2\sin^2x)}\nonumber\\
&=\frac{K_D}{2}\int_{-\infty}^{\infty}dx\frac{e^{-x^2\alpha-i
x\beta}(x^2-\sin^2x)}{x^2(x^2-v_D^2\sin^2x)}\;.
\end{align}
Eq.~(\ref{PDas-3}) can be approximately evaluated by the saddle
point $x=-i\beta/(2\alpha)$. With the help of the identity for a
saddle point approximation at $x=x_0$
\begin{equation}
\int_{-\infty}^{\infty}dx
e^{-\alpha{f(x)}}g(x)\approx\sqrt{\frac{2\pi}{|\alpha{f''}(x_0)|}}e^{-\alpha{f(x_0)}}g(x_0)\;,
\end{equation}
 we have
\begin{equation}
I\approx\frac{a_D^2\sigma^2\lambda^2e^{-\sigma^2\Omega^2}}{24\pi}.
\end{equation}
Here, we have used the relation
$|\omega|=a_D/(\gamma_D\sqrt{\gamma_D^2-1})$, $\gamma_D\gg1$ and
$1\gg{a}_D|\Omega|\sigma^2$ so as to approximate the  result.

 Therefore, for $\gamma_D\gg a_D\sigma$, $\gamma_D\gg1\gg{a}_D|\Omega|\sigma^2$, the transition probabilities take
following approximate form
\begin{equation}
P_D\approx\frac{a_D^2\sigma^2\lambda^2e^{-\sigma^2\Omega^2}}{24\pi}+\frac{\lambda^2}{4\pi}\Big[e^{-\Omega^2\sigma^2}-\sqrt{\pi}\Omega
\sigma {\rm{Erfc}}\big(\Omega\sigma\big)\Big]\;.
\end{equation}

\subsection{approximate forms of $T_{\rm{EDR}}$}\label{approx2}
 In the case of  the infinitely long interaction time and high speed
($\sigma\rightarrow\infty$,$v_D\rightarrow1$), the Gaussian
switching functions can be ignored, the EDR temperature of circular
acceleration can be obtained by substituting Eq.~(\ref{PDas-re})
into Eq.~(\ref{Tedgs}). In the assumption of $a_D\ll|\Omega|$, it is
straightforward to get the EDR temperature $T_{\mathrm{EDR}}\approx
{a}_D/2\sqrt{3}$. This result can also be obtained by
straightforwardly calculating the transition probabilities per unit
proper time~\cite{Bell:1983,Audretsch :1995}. Similarly, the EDR
temperature for linear acceleration  in the case of the infinitely
long interaction time ($\sigma\rightarrow\infty$ ) is exactly equal
to $a_D/(2\pi)$~\cite{Bell:1983,Audretsch :1995,Birrell:1984}.

If the duration time is finite, for vanishingly small speed and
large acceleration with a not-extremely small energy gap
($1\gg{v}_D\;,a_D\sigma\gg|\Omega|\sigma>1\;$), the integrand in the
first term of Eq.~(\ref{PAPB}) can be approximately written as
\begin{equation}
\frac{\cos(x\beta)e^{-x^2\alpha}(x^2-\sin^2x)}{x^2(x^2-v_D^2\sin^2x)}\approx\frac{\cos(x\beta)e^{-x^2\alpha}(x^2-\sin^2x)}{x^4}\;.
\end{equation}
Because of $1\gg{v}_D$ and $a_D\gg|\Omega|$,  Eq.~(\ref{PDas-alph})
 tells us that $\alpha\ll1$ and $\beta\ll1$.
Thus, the vanishingly small $\alpha$ and $\beta$ lead to
\begin{align}
&K_D\int_{0}^{\infty}dx\frac{\cos(x\beta)e^{-x^2\alpha}(x^2-\sin^2x)}{x^4}\nonumber\\&\approx
K_D\int_{0}^{\infty}dx\frac{(x^2-\sin^2x)}{x^4}=\frac{K_D\pi}{3}\;.
\end{align}
 Assuming the energy gap is not-extremely small
($|\Omega|\sigma>1$),  the second term in Eq.~(\ref{PAPB}) can be
approximately estimated as
\begin{equation}
\frac{\lambda^2}{4\pi}\Big[e^{-\Omega^2\sigma^2}-\sqrt{\pi}\Omega
\sigma {\rm{Erfc}}\big(\Omega\sigma\big)\Big]\approx\left\{ \begin{aligned}
       & 0\;,\quad &  \quad\Omega>0\;\\
          &-\frac{\lambda^2\sigma\Omega}{2\sqrt{\pi}}\;,\quad &\quad\Omega<0\;.
                          \end{aligned} \right.
\end{equation}
Therefore, for a positive $\Omega$, we have
\begin{equation}\label{Tas-1}
P_D(\Omega)\approx\frac{K_D\pi}{3}\;,\quad\quad\;P_D(-\Omega)\approx\frac{K_D\pi}{3}+\frac{\lambda^2\sigma\Omega}{2\sqrt{\pi}}\;.
\end{equation}
Substituting  Eq.~(\ref{Tas-1})  into Eq.~(\ref{Tedgs}), and
applying $|\omega|=a_D(1-{v_D}^2)/v_D$ and $a_D\gg|\Omega|$, it is
easy to get $T_{\mathrm{EDR}} \approx a_D v_D \sqrt{1 - v_D^2}/ 6$
for circular acceleration.

Similarly, in regard to uniformly accelerated motion,  the first term in Eq.~(\ref{PApB-UA}) can be approximately evaluated, for an extremely large  acceleration ($a_D\sigma\gg|\Omega|\sigma>1$), as
\begin{align}
&\frac{\lambda^2a_D\sigma}{4\pi^{3/2}}\int_{0}^{\infty}dx
\frac{(\sinh^2x-x^2)\cos(2x\Omega/{a_D})}{x^2\sinh^2x}{e}^{-{x^2}/{(a_D^2\sigma^2)}}\nonumber\\&\approx
\frac{\lambda^2a_D\sigma}{4\pi^{3/2}}\int_{0}^{\infty}dx
\frac{\sinh^2x-x^2}{x^2\sinh^2x}=\frac{\lambda^2a_D\sigma}{4\pi^{3/2}}\;.
\end{align}
Therefore, for a positive $\Omega$,
\begin{equation}\label{Tas-2}
P_D^{\rm{UA}}(\Omega)\approx\frac{\lambda^2a_D\sigma}{4\pi^{3/2}}\;,\quad\quad\;P_D^{\rm{UA}}(-\Omega)\approx\frac{\lambda^2a_D\sigma}{4\pi^{3/2}}+\frac{\lambda^2\sigma\Omega}{2\sqrt{\pi}}\;.
\end{equation}
Thus, the EDR temperature for uniform acceleration can be expressed as
\begin{equation}
T_{\mathrm{EDR}} \approx \frac{a_D+\pi|\Omega|}{2\pi}\approx\frac{a_D}{2\pi}\;.
\end{equation}

%%%%%%%%%%%%%%%%%%%%%%%%%%%%%%%%%%


\begin{thebibliography}{00}

\bibitem{Plenio:1998}
M. B. Plenio and V. Vedral,
%``Entanglement in quantum information theory,"
 Contemp. Phys. {\bf39}, 431 (1998).
 %[quant-ph/9804075]

\bibitem{Horodecki:2001}
M. Horodecki,
% `` Entanglement measures,"
Quantum Inf. Comput. {\bf 1}, 3 (2001).

\bibitem{Braun:2002}
D. Braun,
%``Creation of entanglement by interaction with a common heat bath,"
Phys. Rev. Lett. {\bf89}, 277901 (2002).
%[arXiv:quant-ph/0205019]

\bibitem{MSkim:2002}
M. S. Kim, J. Lee, D. Ahn, and P. L. Knight,
%``Entanglement induced by a single-mode heat environment,"
Phys. Rev. A {\bf65}, 040101(R) (2002).
%[arXiv:quant-ph/0109052]

\bibitem{Schneider:2002}
S. Schneider and G. J. Milburn,
%`` Entanglement in the steady state of a collective-angular- momentum (Dicke) model,"
 Phys. Rev. A {\bf65},042107 (2002).

\bibitem{Basharov:2002}
A. M. Basharov,
%``Decoherence and entanglement in radiative decay of a diatomic system,"
 J. Exp. Theor. Phys. {\bf94}, 1070 (2002).

\bibitem{Jakobczyk:2002}
 L. Jak\'{o}bczyk,
 %``Entangling two qubits by dissipation, "
 J. Phys. A {\bf35}, 6383 (2002).

\bibitem{Reznik:2003}
B. Reznik,
%``Entanglement from the Vacuum,"
 Found. Phys. {\bf33}, 167 (2003).
 %[arXiv:quant-ph/0212044]

\bibitem{Benatti:2003}
F. Benatti, R. Floreanini and M. Piani,
%``Environment induced entanglement in Markovian dissipative dynamics, "
Phys. Rev. Lett. {\bf91}, 070402 (2003).
%[arXiv:quant-ph/0307052]

\bibitem{Ficek:2003}
Z. Ficek and R. Tana\'{s},
%``Entanglement induced by spontaneous emission in spatially extended two-atom systems,"
 J. Mod. Opt. {\bf50}, 2765 (2003).
 %[arXiv:quant-ph/0302124]

\bibitem{TYu:2004}
T. Yu and J. H. Eberly,
 %``Finite-time disentanglement via spontaneous emission, "
 Phys. Rev. Lett. {\bf93}, 140404 (2004).
 %[arXiv:quant-ph/0404161]

\bibitem{Eberly:2007}
J. H. Eberly and T. Yu,
%``The end of an entanglement, "
Science {\bf316},555 (2007).

\bibitem{Ficek:2006}
Z. Ficek and R. Tana\'{s},
%``Dark periods and revivals of entanglement in a two-qubit system,"
 Phys. Rev. A {\bf74}, 024304 (2006).
 %[arXiv:quant-ph/0604053]

\bibitem{Valentini:1991}
A. Valentini,
%``Non-local correlations in quantum electrodynamics,"
Phys. Lett. A {\bf153},321 (1991).

\bibitem{Ver Steeg:2009}
G.L. Ver Steeg and N.C. Menicucci,
%`` Entangling power of an expanding universe,"
Phys. Rev. D {\bf79}, 044027 (2009).
%[arXiv:0711.3066 [quant-ph]]

\bibitem{Olson:2011}
S.J. Olson and T.C. Ralph,
%``Entanglement between the future and past in the quantum vacuum,"
 Phys. Rev. Lett. {\bf106},  110404 (2011).
%[arXiv:1003.0720[quant-ph]]

\bibitem{BLHu:2012}
B.L. Hu, S.-Y. Lin and J. Louko,
%``Relativistic quantum information in detectors-field interactions,"
 Class. Quant. Grav. {\bf29}, 224005 (2012).
% [arXiv:1205.1328[quant-ph]]

\bibitem{Pozas-Kerstjens:2015}
 A. Pozas-Kerstjens and E. Mart\'{i}n-Mart\'{i}nez,
% Harvesting correlations from the quantum vacuum,
Phys. Rev. D {\bf92},064042 (2015).
%[arXiv:1506.03081[quant-ph]]

\bibitem{EDU:2016-1}
E. Mart\'{i}n-Mart\'{i}nez, A.R.H. Smith and D.R. Terno,
%``Spacetime structure and vacuum entanglement,"
 Phys. Rev. D {\bf93}, 044001 (2016).
%[arXiv:1507.02688[quant-ph]]

\bibitem{EDU:2016-2}
E. Mart\'{i}n-Mart\'{i}nez and B.C. Sanders,
%``Precise space-time positioning for entanglement harvesting, "
New J. Phys. {\bf18}, 043031 (2016).
%[arXiv:1508.01209[quant-ph]]

\bibitem{Pozas-Kerstjens:2016}
A. Pozas-Kerstjens and E. Mart\'{i}n-Mart\'{i}nez,
%Entanglement harvesting from the electromagnetic vacuum with hydrogenlike atoms,"
 Phys. Rev.D {\bf94}, 064074 (2016).
 %[arXiv:1605.07180[quant-ph]]

\bibitem{Zhjl:2018}
L.J. Henderson, R.A. Hennigar, R.B. Mann, A.R.H. Smith and J. Zhang,
%``Harvesting entanglement from the black hole vacuum,"
 Class. Quant. Grav. {\bf35}, 21LT02 (2018).
% [arXiv:1712.10018[quant-ph]]

\bibitem{Zhjl:2019}
L.J. Henderson, R.A. Hennigar, R.B. Mann, A.R.H. Smith and J. Zhang,
%``Entangling detectors in anti-de Sitter space, "
JHEP {\bf05},178 (2019).
%[arXiv:1809.06862[quant-ph]]

\bibitem{Ng:2018}
K. K. Ng, R. B. Mann, and E. Mart\'{i}n-Mart\'{i}nez,
%``New techniques for entanglement harvesting in flat and curved spacetimes,"
 Phys. Rev. D {\bf97}, 125011 (2018).
 %[arXiv:1805.1096[quant-ph]]

\bibitem{Ng:2018-2}
K. K. Ng, R. B. Mann and E. Mart\'{i}n-Mart\'{i}nez,
%``Unruh-DeWitt detectors and entanglement: the anti-de Sitter space,"
 Phys. Rev. D {\bf98},125005 (2018).
%[arXiv:1809.06878[quant-ph]]

\bibitem{Salton:2015}
G. Salton, R. B. Mann, and N. C. Menicucci,
%`` Acceleration-assisted entanglement harvesting and rangefinding,"
 New J. Phys.{\bf17},035001 (2015).
 %[arXiv:1408.1395[quant-ph]]

 \bibitem{Unruh:1976}
 W. G. Unruh,
 %`` Notes on black-hole evaporation,"
 Phys. Rev. D {\bf14}, 870 (1976).

 \bibitem{Birrell:1984}
  N. D. Birrell and P. C. W. Davies, Quantum Fields in
Curved Space, Cambridge Monographs on Mathematical Physics (Cambridge Univ. Press, Cambridge, UK, 1984).

\bibitem{Crispino:2008}
L. C. B. Crispino, A. Higuchi, G. E. A. Matsas,
%`` The Unruh effect and its applications,"
Rev.Mod. Phys. {\bf80}, 787 (2008).
%[arXiv:0710.5373[gr-qc]]

\bibitem{EDU:2011}
E. Mart\'{i}n-Mart\'{i}nez, I. Fuentes, Robert B. Mann,
%``Using Berry¡¯s Phase to Detect the Unruh Effect at Lower Accelerations,"
 Phys. Rev. Lett. {\bf107}, 131301(2011).
 %[arXiv:1012.2208[quant-ph]]

\bibitem{Hu:2012}
J. Hu, H. Yu,
%``Geometric phase for an accelerated two-level atom and the Unruh effect,"
Phys. Rev. A {\bf85},  032105 (2012).
%[arXiv:1203.5869[quant-ph]]

\bibitem{Zhjl:2016}
H. Zhai., J. Zhang and H. Yu,
%``Geometric phase of an accelerated two-level atom in the presence of a perfectly reflecting plane boundary,"
 Ann. Phys. (N. Y.) {\bf371}, 338 (2016).

 \bibitem{Audretsch :1995}
 J. Audretsch and R. M\"{u}ller,
 %``Radiative energy shifts of an accelerated two-level system,"
  Phys. Rev. A {\bf52}, 629 (1995). % [arXiv:gr-qc/9503058]

 \bibitem{Passante:1998}
 R. Passante,
 %``Radiative level shifts of an accelerated hydrogen atom and the Unruh effect in quantum electrodynamics,'
 Phys.Rev. A {\bf57},1590 (1998).

 \bibitem{Rizzuto:2009}
 L. Rizzuto and S. Spagnolo,
 %'' Energy level shifts of a uniformly accelerated atom in the presence of boundary conditions,"
  J. Phys.: Conf. Ser. {\bf161}, 012031 (2009).

 \bibitem{Zhu:2010}
Z. Zhu and H. Yu,
%``Position-dependent energy-level shifts of an accelerated atom in the presence of a boundary,"
 Phys.Rev. A  {\bf82}, 042108 (2010).

\bibitem{Benatti:2004}
F. Benatti and R. Floreanini,
%``Entanglement generation in uniformly accelerating atoms: Re- examination of the Unruh effect,"
 Phys. Rev.A {\bf70}, 012112 (2004).

\bibitem{Zhjl:2007}
 J. Zhang and H. Yu,
 %``Unruh effect and entanglement generation for accelerated atoms near arefecting boundary,"
  Phys. Rev. D {\bf75}, 104014 (2007).
 %[arXiv:0705.1092 [gr-qc]]

\bibitem{Landulfo:2009}
A. G. S. Landulfo and G. E. A. Matsas,
%``Sudden death of entanglement and teleportation fidelity loss via the Unruh effect,"
 Phys. Rev. A {\bf80}, 032315 (2009).
 %[arXiv:0907.0485 [gr-qc]]

\bibitem{Doukas:2010}
J. Doukas and B. Carson,
%``Entanglement of two qubits in a relativistic orbit,"
 Phys. Rev. A{\bf 81}, 062320 (2010).
 %[arXiv:1003.2201[quant-ph]]

\bibitem{Ostapchuk:2012}
 D. C. M. Ostapchuk, S.-Y. Lin, R. B. Mann and B. L. Hu,
% ``Entanglement dynamics between inertial and non-uniformly accelerated detectors,"
 JHEP {\bf07}, 072 (2012).
 %[arXiv:1108.3377 [gr-qc]]

\bibitem{Hu:2015}
 J. Hu and H. Yu, %``Entanglement dynamics for uniformly accelerated two-level atoms,"
  Phys.Rev. A {\bf91}, 012327 (2015).
  %[arXiv:1501.03321[quant-ph]]
\bibitem{Cheng:2018}
 S. Cheng, H. Yu, and J. Hu,
 %`` Entanglement dynamics for uniformly accelerated two-level atoms in the presence of a reflecting boundary,"
  Phys. Rev. D {\bf98}, 025001 (2018).
  %[arXiv:1806.05344[quant-ph]]

\bibitem{Koga:2019}
J. I. Koga, K. Maeda, G. Kimura,
%`` Entanglement extracted from vacuum into accelerated Unruh-DeWitt detectors and energy conservation,
Phys. Rev. D {\bf100}, 065013 (2019).
%arXiv:1906.02843[quant-ph]

\bibitem{She:2019}
J. She, J. Hu, H. Yu,
%``Entanglement dynamics for circularly accelerated two-level atoms coupled with electromagnetic vacuum fluctuations,"
Phys. Rev. D {\bf99}, 105009 (2019).
%[arXiv:1904.10111[quant-ph]]

\bibitem{Brenna:2016}
W. G. Brenna, R. B. Mann, and E. Martin-Martinez,
%``Anti-Unruh Phenomena,"
Phys. Lett. B {\bf757}, 307 (2016).
% arXiv:1504.02468 [quantph].

\bibitem{Garay:2016}
L. J. Garay, E. Martin-Martinez, and J. de Ramon,
%``Thermalization of particle detectors: The Unruh effect and its reverse,"
Phys. Rev. D {\bf94}, 104048 (2016).
%arXiv:1607.05287 [quant-ph].

\bibitem{Li:2018}
T. Li, B. Zhang, L. You, %``Would quantum entanglement be increased by anti-Unruh effect?"
 Phys.Rev.D {\bf97}, 045005 (2018).
 %arXiv:1802.07886 [gr-qc]


\bibitem{Bell:1983}
J. S.Bell and J. M. Leinaas,
%`` electrons as accelerated thermometers,"
Nucl. Phys. B {\bf212}, 131 (1983).

\bibitem{DeWitt:1979}
 B. S. DeWitt, S. Hawking, and W. Israel, General Relativity: An Einstein Centenary Survey (Cambridge University
Press Cambridge, 1979).

 \bibitem{Wootters:1998}
 W. K. Wootters,
 %``Entanglement of Formation of an Arbitrary State of Two Qubits,"
  Phys. Rev. Lett. {\bf80}, 2245 (1998).

\bibitem{Kim:1987}
S. K. Kim,K. S. Soh,J. H. Yee,
%``Zero-point field in a circular-motion frame,"
Phys. Rev. D{\bf 35},557 (1987).

\bibitem{Kubo:1957}
R. Kubo,
% ¡°Statistical-mechanical theory irreversible processes. I.general theory and simple applications to magnetic and conduction problems,¡±
J. Phys. Soc. Jpn. {\bf12}, 570 (1957).

\bibitem{Martin:1959}
P. C. Martin and J. S. Schwinger,
%``Theory of Many-Particle Systems. I,"
Phys. Rev. {\bf115}, 1342 (1959).

\bibitem{Nambu:2013}
Y. Nambu,
%``Entanglement Structure in Expanding Universes, "
Entropy {\bf15},1847-1874 (2013).
%arXiv:1305.4193 [gr-qc]

\bibitem{Fewster:2016}
C. J. Fewster, B. A. Jurez-Aubry and J. Louko,
%``Waiting for Unruh,"
 Class.Quant. Grav. {\bf33}, 165003 (2016).
%arXiv:1605.01316 [grqc].

\bibitem{Bogolubov:1990}
N. N. Bogolubov, A. A. Logunov, A. I. Oksak, and I. T.
Todorov, General Principles of Quantum Field Theory
(Springer, 1990).

\end{thebibliography}
\end{document}